\documentclass[a4paper,11pt]{article}
\usepackage{jcappub} % for details on the use of the package, please see the JINST-author-manual
\usepackage{lineno}
\usepackage{array} % For better column control
\usepackage{booktabs} % For professional-looking tables
\renewcommand{\arraystretch}{1.3}
\usepackage{hyperref}
\hypersetup{colorlinks=true,linkcolor=blue,citecolor=cyan}
\usepackage{graphicx}
\usepackage{subcaption}
\usepackage{xcolor, soul}
\sethlcolor{green}
\usepackage{dcolumn}
\usepackage{bm}
\usepackage{color}
\usepackage{enumitem}
\usepackage{mathrsfs}
\usepackage{amsmath}
\usepackage{amssymb}
\usepackage{cleveref}
\usepackage{orcidlink}

\crefname{equation}{Eqn.}{Eqns.}
\crefname{figure}{Fig.}{Figs.}
\crefname{section}{Sec.}{Sec.}
\crefname{table}{Table}{Tables}
\usepackage{physics}
\usepackage{multirow}
\setstcolor{red}
\arxivnumber{2603.17986} % Only if you have one
\title{\textcolor{black}{Observational Signatures of Exact Black Hole Solutions in a Dark Matter Halo}}

\author[a]{Azalbek Boltaev}
\affiliation[a]{New Uzbekistan University, Movarounnahr str. 1, Tashkent 100000, Uzbekistan} 

\author[b,c,d]{Tursunali Xamidov}
\affiliation[b]{Institute for Theoretical Physics and Cosmology,
Zhejiang University of Technology, Hangzhou 310023, China}
\affiliation[c]{Institute of Fundamental and Applied Research, National Research University TIIAME, Kori Niyoziy 39, Tashkent 100000, Uzbekistan} 

\affiliation[d]{Tashkent State Technical University, 100095 Tashkent, Uzbekistan}

\author[c,e,f]{Sanjar Shaymatov}
\affiliation[e]{University of Tashkent for Applied Sciences, Str. Gavhar 1, Tashkent 100149, Uzbekistan}
\affiliation[f]{National University of Uzbekistan, Tashkent 100174, Uzbekistan}

\emailAdd{azalbekboltaev@gmail.com}
\emailAdd{xamidovtursunali@gmail.com}
\emailAdd{sanjar@astrin.uz}

 \abstract{
In this work, we derive novel exact solutions describing Schwarzschild-like black holes (BHs) embedded in a Dehnen-type dark matter (DM) halo density profile and investigate their geometric, dynamical, and observational signatures arising from such geometries. We begin by analyzing the horizon structure and spacetime curvature invariants, as well as examining the energy conditions associated with the DM halo. Subsequently, we study the influence of the DM halo on both timelike and null geodesics in the resulting geometry. 
\textcolor{black}{
Finally, we obtain observational constraints on the DM halo parameters by comparing the model predictions with weak-field data from Mercury and the S2 star orbit, as well as with the Event Horizon Telescope (EHT) shadow measurements of M87$^\star$ and Sgr~A$^\star$, combined with the corresponding EHT and GRAVITY measurements of the BH mass and distance. Using Bayesian inference and Markov Chain Monte Carlo (MCMC) methods, we obtain the posterior distributions of the model parameters and derive 95\% upper limits on the DM halo parameters. For M87$^\star$, we obtain $\rho_s<11.83~M_\odot\,{\rm pc}^{-3}$ and $r_s<54.36~{\rm kpc}$. For Sgr~A$^\star$, the EHT, GRAVITY, and combined EHT+GRAVITY analyses give $\rho_s<31.56$, $16.76$, and $21.25~M_\odot\,{\rm pc}^{-3}$, respectively. We find that all three cases lead to similar constraints on $r_s$ for Sgr~A$^\star$. 
}
Our analysis provides valuable insight into probing the potential influence of DM halo environments on spacetime geometry and observable properties of astrophysical BHs, offering an alternative perspective on BH–DM interactions.
}

\def\cred#1{\textcolor{black}{#1}}

\begin{document}
\maketitle
\flushbottom

%%%%%%%%%%%%%%%%%%%%%%
\section{Introduction}
\label{introduction}
%%%%%%%%%%%%%%%%%%%%%%

Despite numerous experimental tests in both the weak- and strong-field regimes, General Relativity (GR) still faces significant challenges and unresolved issues. In particular, it does not provide a satisfactory description of black hole (BH) singularities, nor does it explain the unknown nature of dark matter (DM) and dark energy. Recent observations by the EHT \cite{Akiyama19L1,Akiyama22L12} and LIGO–Virgo \cite{Abbott16a,Abbott16b} have confirmed key predictions of BH physics.  Nevertheless, GR is generally regarded as an incomplete theory. Consequently, BHs continue to be intensively investigated using these remarkable observational advances, and the origin and fundamental nature of DM remain among the major open problems in modern cosmology. In astrophysical scenarios, supermassive BHs located at galactic centers are expected to reside in complex environments. Observational and theoretical studies further suggest that they are embedded within DM halos \cite{Iocco15NatPhy,Bertone18Nature}. Moreover, compelling evidence for DM arises from galactic rotation curves \cite{Rubin70ApJ}, observations of galaxy clusters such as the Bullet Cluster \cite{Clowe06ApJL}, and the formation of large-scale cosmic structures \cite{Davis85ApJ}.

Astrophysical observations indicate that dark matter (DM) accounts for about $\sim$ 90\% of the mass of a galaxy, while the remaining $\sim$ 10\% consists of luminous baryonic matter \cite{Persic96}. Many large spiral and elliptical galaxies are also observed to host central supermassive BHs embedded within these DM halos \cite{Valluri04ApJ,Akiyama19L1,Akiyama19L6,Akiyama22L12}. Extensions of the Standard Model predict several viable DM candidates, such as weakly interacting massive particles (WIMPs), axions, and neutrinos \cite{Boehm04NPB,Bertone05PhR,Feng09JCAP,Schumann19}. In such scenarios, where interactions with Standard Model particles are extremely weak, the properties of DM are primarily probed through its gravitational effects. Hence, DM is expected to accumulate around supermassive BHs at galactic centers, where it may potentially affect the extreme \cite{Babak17PRD} and intermediate \cite{Brown07PRL,Amaro-Seoane18PRD} mass ratio inspirals and provide insights into the DM halo profile. Such halos are also essential for explaining galactic rotation curves and cluster dynamics \cite{Rubin70ApJ,Bertone18Nature,Corbelli00MNRAS,Clowe06ApJL}.

BHs, which are believed to reside within DM environments, provide a natural laboratory for studying BH–DM interactions. Given the fundamental role of DM halos, investigating this interplay and developing viable DM models is essential for improving our understanding of the nature of DM. To this end, BH solutions embedded in DM halos have been modeled using several analytical profiles, including the Einasto, Navarro–Frenk–White, Burkert, and Dehnen profiles \cite{Merritt06ApJ,Dutton_2014,Navarro96ApJ,Burkert95ApJ,Dehnen93,Shukirgaliyev21A&A,Gohain24DM,Pantig22JCAP,Al-Badawi25JCAP,Uktamov25EPJC,Wang2025PhRvD,Gabriel2026arXiv260316402G}. Alternative approaches involve DM distributions associated with phantom scalar fields and analytical supermassive BH solutions within DM halos \cite{Li-Yang12,Shaymatov21d,Shaymatov21pdu,Shaymatov22a,Cardoso22DM,Hou18-dm,Shen24PLB,Shen25PLB, BoWang2025JCAP...01..086L}.
 
It is important to emphasize that a DM halo surrounding a BH not only affects the particle dynamics but also modifies the spacetime geometry. This may alter the horizon structure and leave observable signatures in quantities such as the innermost stable circular orbit (ISCO), BH shadows, etc. Such DM–BH interactions have also been explored in the context of the Dehnen-type halo framework. For example, BH solutions embedded in Dehnen-type DM halos have been proposed to model ultra-faint dwarf galaxies \cite{Pantig22JCAP}. Subsequent studies have further explored BH solutions with different DM halo density profiles \cite{Gohain24DM,Al-Badawi25JCAP,Uktamov25EPJC}, investigating their effects on quasinormal modes, BH shadows, and gravitational waveforms, as well as constraining DM halo parameters \cite{Al-Badawi25CPC,Al-Badawi25CTP_DM,Alloqulov-Xamidov25,Xamidov25PDU,Xamidov25EPJC...85.1193X,BoWang2026Univ...12...48C}.
\cred{
More specifically, previous studies of Dehnen-halo BH models considered $(\alpha,\beta,\gamma)=(1,4,5/2)$ in Refs.~\cite{Al-Badawi25JCAP,Al-Badawi25CTP_DM}, while Refs.~\cite{Uktamov25EPJC} and \cite{Gohain24DM} studied $(1,4,2)$ and $(1,4,0)$, respectively. These studies therefore focused on particular values of $\gamma$. In the present work, we consider the general $(1,4,\gamma)$ family with $0\leq\gamma<3$, keeping the inner-slope parameter $\gamma$ free. In addition, unlike the previous constructions based on a Newtonian approximation for the DM contribution, we obtain the metric function directly from the Einstein field equations for an effective anisotropic matter source with a Dehnen density profile. In this context, we present Schwarzschild-like BH solutions in a Dehnen-type DM halo with a density profile $(1,4,\gamma)$.
}

We analyze spacetime curvature invariants, examine the associated energy conditions, and investigate the influence of the DM halo on timelike and null geodesics. 

\cred{
Constraints are also obtained from the EHT shadow measurements of M87$^\star$ and Sgr~A$^\star$, together with the corresponding EHT and GRAVITY measurements of the black-hole mass and distance. Bayesian inference and Markov Chain Monte Carlo (MCMC) methods are employed to estimate the model parameters.
}

The paper is organized as follows: In Sec.~\ref{Sec:II}, we present exact solutions of the Einstein field equations describing Schwarzschild-like BHs embedded in a DM halo characterized by a Dehnen-type density profile, and analyze their horizon structure. In Sec.~\ref{Sec:III}, we examine the spacetime curvature invariants and investigate the energy conditions associated with the DM halo. We then study the influence of the DM halo on both timelike and null geodesics in the resulting geometry for the density profile $(1, 4, \gamma)$. In Sec.~\ref{Sec:IV}, we constrain the characteristic DM halo parameters using weak-field observational data from Mercury and the orbit of the S2 star. 
\cred{
In Sec.~\ref{Sec:V}, we use the EHT shadow measurements of M87$^\star$ and Sgr~A$^\star$, together with the corresponding EHT and GRAVITY mass and distance measurements, to constrain the halo parameters.
}
Using Bayesian inference and MCMC methods, we obtain constraints on the characteristic density $\rho_s$ and scale radius $r_s$ of the DM halo. Finally, we summarize our results and conclusions in Sec.~\ref{Sec:con}.

%%%%%%%%%%%%%%%%%%%%%
\section{Schwarzchild-like black hole spacetime with the dark matter halo}\label{Sec:II}
\begin{figure*}[ht!]
\centering
\includegraphics[width=0.45\textwidth]{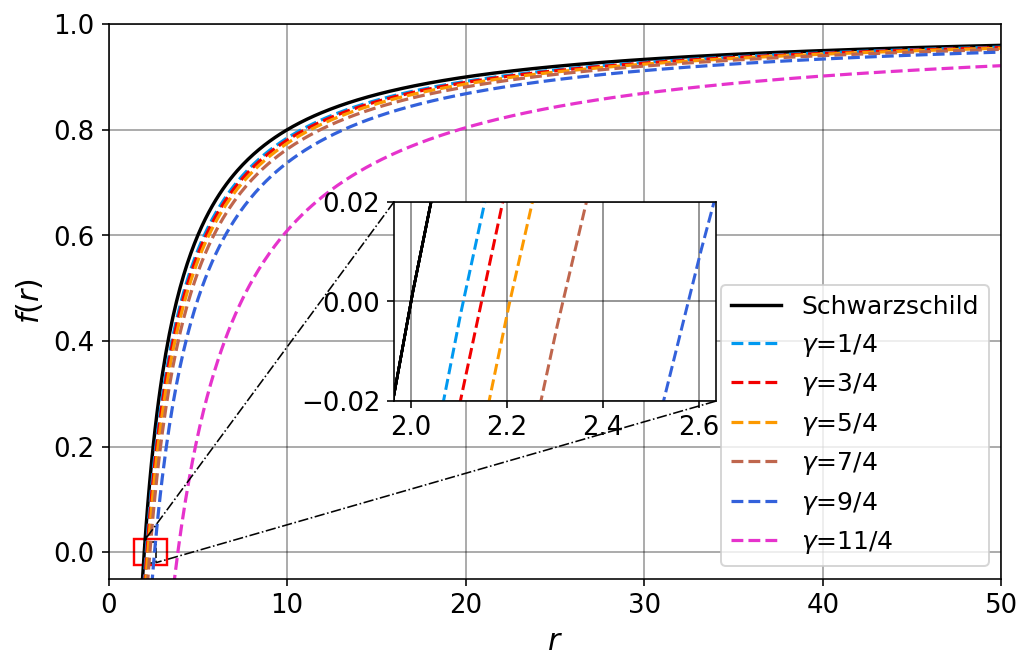}
\includegraphics[width=0.44\textwidth]{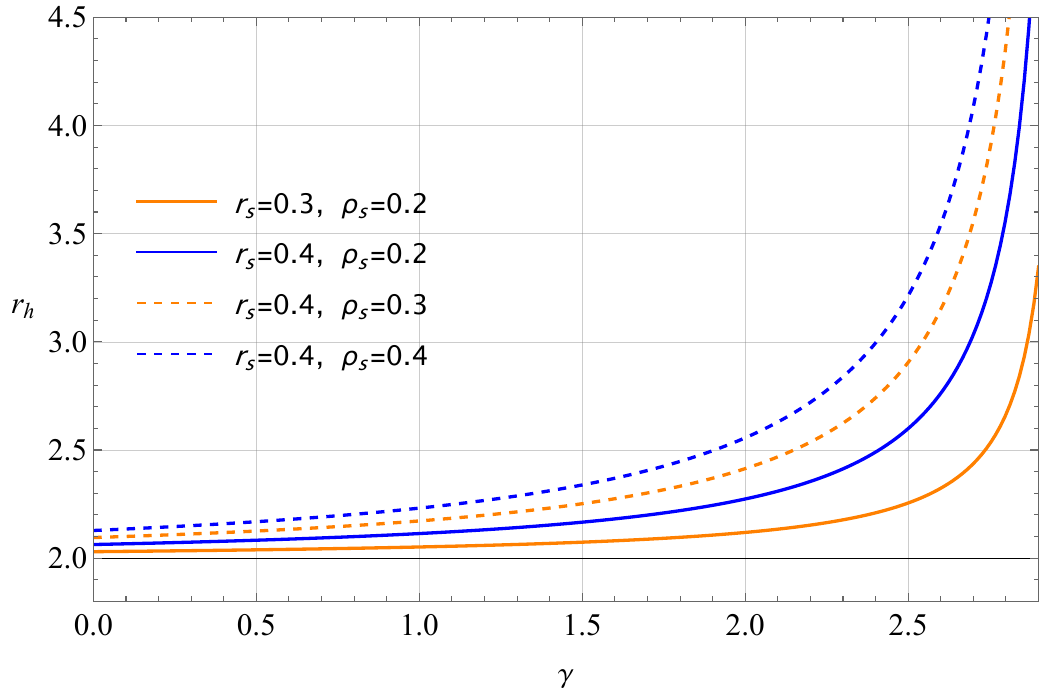}
\caption{\centering Left panel: The radial dependence of function $f(r)$ for various values of parameter $\gamma$ for the fixed values of the DM halo parameters $r_s=0.4$ and $\rho_s=0.3$. Right panel: The dependence of horizon radius on $\gamma$ for various combinations of $r_s$ and $\rho_s$.
\label{Fig.f(r)}}
\end{figure*}
Here, we begin by considering a spherically symmetric background spacetime with a DM halo distribution, modeling through a Dehnen-type density profile. With this in mind, we derive an exact solution of the Einstein field equations describing a Schwarzschild-like BH embedded in a Dehnen-type DM halo mass distribution \cite{Mo10book}. To this end, we first specify the form of the density profile that yields corresponding Dehnen-type DM halo mass distribution. In general, it can be written as   
\begin{eqnarray}\label{eq.density}
    \rho(r)=\rho_s\left(\frac{r}{r_s}\right)^{-\gamma}\Big[\left(\frac{r}{r_s}\right)^\alpha+1\Big]^{\frac{\gamma-\beta}{\alpha}}\, ,
\end{eqnarray}
with the characteristic density $\rho_s$ and characteristic scale $r_s$  of the DM halo. In addition to these characteristic DM halo parameters, $\alpha$, $\beta$ and $\gamma$ determine the specific shape of the density profile, referred to as the free parameters of the profile. For instance, the parameter $\gamma$ is restricted to the interval $0\leq\gamma<3$. Taking this into account, in this work, we choose the following specific profile parameters  ($\alpha,\beta,\gamma)=(1, 4, \gamma)$. Following Eq.~(\ref{eq.density}), we then define the DM halo mass profile as
\begin{eqnarray}\label{eq.mass}
    M_D(r)=\int_0^r 4\pi\rho(r_{\mathrm{_1}})r^2_{\mathrm{_1}} dr_{\mathrm{_1}}\, .
\end{eqnarray}

We then turn to write the line elements of a static and spherically symmetric spacetime surrounded by the DM halo, modeling through redshift $A(r)$ and shape $B(r)$ functions
\begin{eqnarray}\label{eq.line}
    ds^2=-A(r)dt^2+\frac{dr^2}{B(r)}+r^2d\Omega^2\, ,
\end{eqnarray}
with the solid angle with spherical coordinates $d\Omega^2=d\theta^2+\sin^2{\theta}d\phi^2$. 
For that the Einstein field equation reads as (\cite{Xu18JCAP,Al-Badawi25JCAP}):
\begin{eqnarray}\label{eq.Einstein 1}
    R_{\mu\nu}-\frac{1}{2}g_{\mu\nu}R=8\pi T_{\mu\nu}(D)\, ,
\end{eqnarray}
where $T_{\mu\nu}(D)$ refers to the source part. It is then given by    
\begin{eqnarray}
T_\mu^\nu(D)=g^{\nu\alpha}T_{\mu\alpha}(D)=\text{diag}[-\rho(r),P_r(r),P_t(r),P_t(r)]\, ,\nonumber
\end{eqnarray} 
referred to as the energy-momentum tensor of the DM halo spacetime geometry. 
Given the energy-momentum tensor, we write the Einstein's field equations as follows:
\begin{eqnarray}\label{eq.temporal}
   B(r)\left(\frac{1}{r}\frac{B'(r)}{B(r)}+\frac{1}{r^2}\right)-\frac{1}{r^2}= 8\pi T_t^t(D)\, ,
\end{eqnarray}
%%%%%%%%%%%%%
\begin{eqnarray}\label{eq.radial}
    B(r)\left(\frac{1}{r}\frac{A'(r)}{A(r)}+\frac{1}{r^2}\right)-\frac{1}{r^2}=8\pi T_r^r(D)\, ,
\end{eqnarray}
%%%%%%%%%%%%%
\begin{eqnarray}\label{eq.theta}
    \;&\frac{B(r)}{2 A(r)}\left(A''(r)-\frac{A'(r)^2}{2 A(r)}+\frac{A'(r)}{r}\right)+
    \frac{B'(r)}{2} \left(\frac{A'(r)}{2 A(r)}+\frac{1}{r}\right)\nonumber\\
    \;&\quad=8\pi T_\theta^\theta(D)=8\pi T_\phi^\phi(D)=8\pi P_t(r)\, .
\end{eqnarray}
%%%%%%%%%%%%
From the temporal component of Einstein’s equations, Eq.~\eqref{eq.temporal}, it is the first-order differential equation with respect to $B'(r)$ and is rewritten as
\begin{eqnarray}
    \frac{d}{dr}\Big[r\big(B(r)-1\big)\Big]=-8\pi r^2 \rho(r)\, , 
\end{eqnarray}
which solves to give 
\begin{eqnarray}\label{eq.B(r)func}
    B(r)=1-\frac{8\pi}{r}\int r^2\rho(r)dr=\nonumber\\
    1-\frac{8\pi\rho_sr_s^3}{(3-\gamma)r}\Big(\frac{r}{r+r_s}\Big)^{3-\gamma}+\frac{C}{r}\, ,
\end{eqnarray}
where $C$ is an integration constant, taken as $-2M$ provided that Eq.\eqref{eq.B(r)func} reduces to the Schwarzschild when the DM halo is absent. 

\cred{
We impose the condition $P_r(r)=-\rho(r)$, which leads to $A(r)=B(r)$. We note that this relation does not describe ordinary collisionless cold dark matter, which would require a phase-space distribution of the particles to determine the radial and tangential pressures. Instead, the source is modeled as an effective anisotropic matter distribution with radial tension, and its energy density follows the Dehnen profile. Within this model, we obtain an exact solution of the Einstein field equations. The resulting spacetime metric can be written as
}

\begin{eqnarray}\label{eq.full-line}
    ds^2=-f(r)dt^2+\frac{1}{f(r)}dr^2+r^{2}\left(
d\theta ^{2}+\sin ^{2}\theta d\phi ^{2}\right)\, , %r^2d\Omega^2\
\end{eqnarray}
where 
\begin{eqnarray}\label{Eq.radial_func}
    f(r)=1-\frac{2M}{r}-\frac{8\pi\rho_sr_s^3}{(3-\gamma)r}\Big(\frac{r}{r+r_s}\Big)^{3-\gamma}\, .
\end{eqnarray}

\cred{
By evaluating the integral in Eq.~\eqref{eq.mass} in the limit of $r\to\infty$, we obtain the mass for the DM halo as
\begin{eqnarray}\label{Eq. total_mass}
    M_{DM}=\frac{4\pi \rho_s r_s^3}{3-\gamma}\, .
\end{eqnarray}
Then, Eq. \eqref{Eq.radial_func} yields as 
\begin{eqnarray} \label{Eq.radial_func_MDM}
    f(r)=1-\frac{2M}{r}-\frac{2M_{DM}}{r}\Big(\frac{r}{r+r_s}\Big)^{3-\gamma}\, .
\end{eqnarray}
Here and throughout this work, $M$ denotes the mass of the
central BH, while the contribution of the DM halo is treated
separately through $\rho_s$ and $r_s$. At large distances,
Eq.~\eqref{Eq.radial_func_MDM} behaves as
\begin{equation}
f(r)=1-\frac{2(M+M_{\rm DM})}{r}+\mathcal{O}(r^{-2}),
\end{equation}
so that the total mass is
$M_{\rm t}=M+M_{\rm DM}$.
}

Using Eq.~\eqref{Eq.radial_func}, we represent a number of the radial functions describing exact BH-DM solutions for specific values of the parameter $\gamma$, as tabulated in Table~\ref{table:functions}.  
%%%%%
\begin{table}[t]
\centering
\renewcommand{\arraystretch}{1.3
}
\setlength{\tabcolsep}{6pt}

\resizebox{0.6\textwidth}{!}{%
\begin{tabular}{|c|c|}
\hline

\textbf{$(\alpha,\beta,\gamma)$} &
$f(r)$
\\

\hline

$(1,4,1/4)$
&
$1-\frac{2M}{r}
-\frac{32\pi\rho_s r_s^3}{11r}
\left(\frac{r}{r+r_s}\right)^{11/4}$
\\
\hline

$(1,4,3/4)$
&
$1-\frac{2M}{r}
-\frac{32\pi\rho_s r_s^3}{9r}
\left(\frac{r}{r+r_s}\right)^{9/4}$
\\
\hline

$(1,4,5/4)$
&
$1-\frac{2M}{r}
-\frac{32\pi\rho_s r_s^3}{7r}
\left(\frac{r}{r+r_s}\right)^{7/4}$
\\
\hline

$(1,4,7/4)$
&
$1-\frac{2M}{r}
-\frac{32\pi\rho_s r_s^3}{5r}
\left(\frac{r}{r+r_s}\right)^{5/4}$
\\
\hline

$(1,4,9/4)$
&
$1-\frac{2M}{r}
-\frac{32\pi\rho_s r_s^3}{3r}
\left(\frac{r}{r+r_s}\right)^{3/4}$
\\
\hline

$(1,4,11/4)$
&
$1-\frac{2M}{r}
-\frac{32\pi\rho_s r_s^3}{r}
\left(\frac{r}{r+r_s}\right)^{1/4}$
\\

\hline
\end{tabular}%
}

\caption{
Exact analytical Schwarzschild-like BH solutions describing static BHs
surrounded by a DM halo with a Dehnen-type density profile
$(1,4,\gamma)$.
}
\label{table:functions}

\end{table}

We now turn to studying the behavior of $f(r)$ and demonstrate its radial dependence in Fig.~\ref{Fig.f(r)}, highlighting how it changes depending on the parameter, $\gamma$, of Dehnen-type DM halo density profile. As can be seen from Fig.~\ref{Fig.f(r)}, the behavior of $f(r)$ reduces to the flat spacetime at larger distances, especially at infinity. Furthermore, a systematic and progressively amplified movement of the curves towards larger $r$-values is observed with the growth of the parameter $\gamma$, resulting in an increase in the strength of the gravitational potential. The spacetime Eq.~(\ref{Eq.radial_func}) has an event horizon, which can be determined by the largest root of $f(r)=0$. As illustrated in the right panel of Fig.~\ref{Fig.f(r)}, the event horizon radius grows rapidly as the parameter $\gamma$ and the DM halo parameters $\rho_s$ and $r_s$ increase.

\section{The spacetime curvature and energy condition characteristics} \label{Sec:III}

\begin{figure*}
    \centering
    \includegraphics[width=0.32\textwidth]{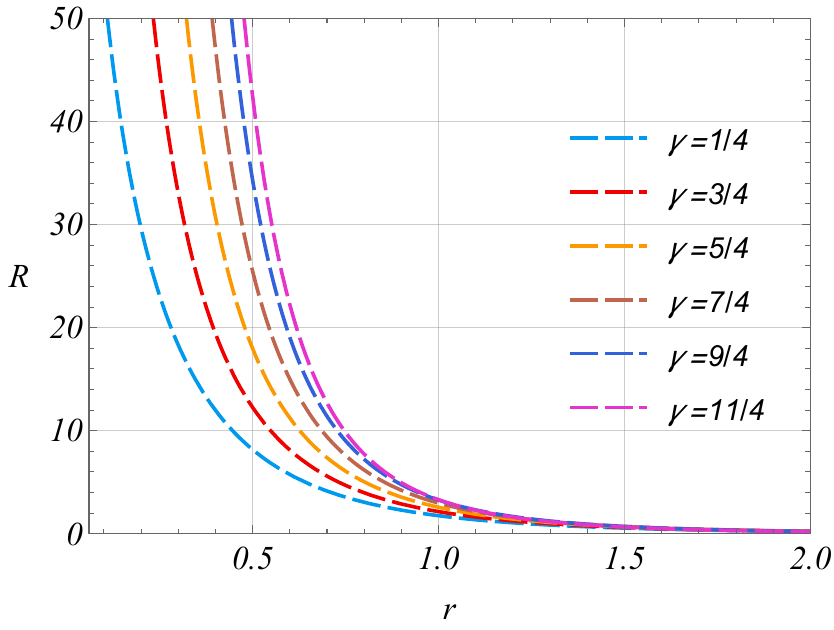}
    \includegraphics[width=0.32\textwidth]{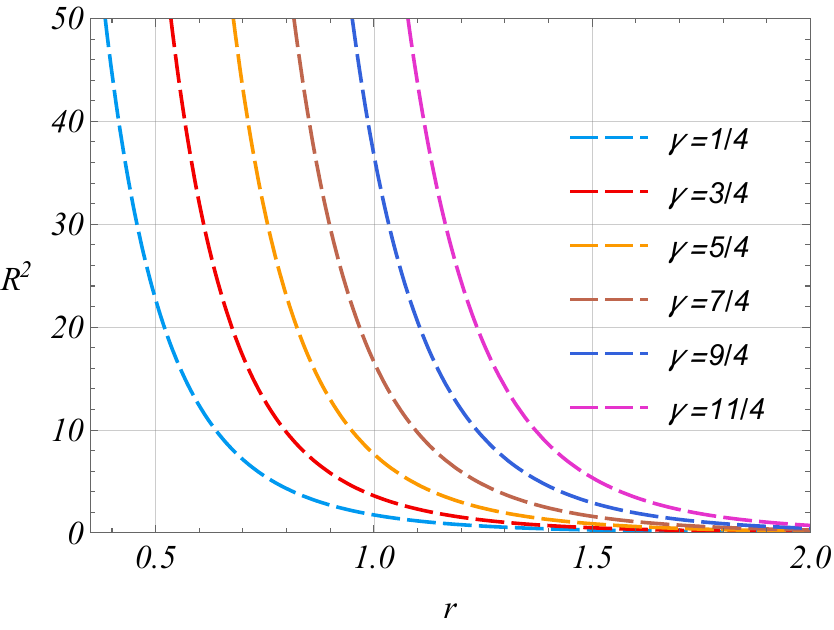}
    \includegraphics[width=0.32\textwidth]{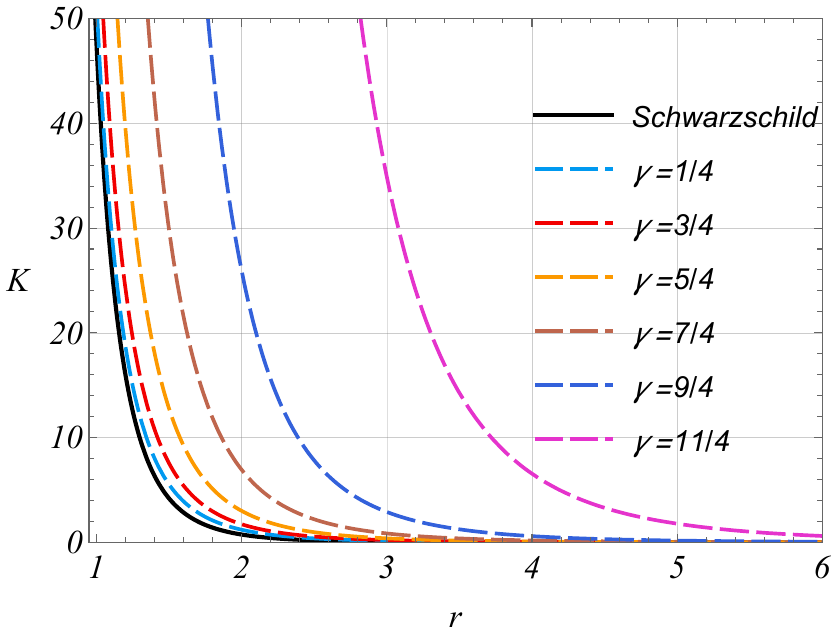}
    \caption{The Ricci scalar($R$) (left panel), the Ricci square ($R^2$) (middle panel), and the Kretschmann scalar (K) (right panel) as a function of r for a Schwarzschild-like BH in a DM halo for different values of $\gamma$ for $r_s=1$ and $\rho_s$=0.5}
    \label{Fig.Curvature}
\end{figure*}

We now investigate the spacetime curvature invariants associated with a modified Schwarzschild-like BH immersed in a DM halo described by a Dehnen-type density profile. To perform this analysis, we compute the principal curvature invariants of the spacetime, namely the Ricci scalar $R$, the Ricci square $R_{\mu\nu}R^{\mu\nu}$, and the Kretschmann scalar $R_{\mu\nu\alpha\beta}R^{\mu\nu\alpha\beta}$. These invariants are analyzed in detail in order to determine the existence of a spacetime singularity at $r=0$. We begin by examining the Ricci scalar for the combined BH–DM configuration, which, in this case, takes the form 
\begin{eqnarray}\label{eq:Ricci}
   R=\frac{8 \pi  (4-\gamma ) r_s^5 \left(\frac{r_s+r}{r}\right)^\gamma \rho _s }{\left(r_s+r\right){}^5}\, .
\end{eqnarray}
When the DM halo is absent (i.e., $\rho _s=0$), the spacetime reduces to the Schwarzschild metric, where the Ricci scalar vanishes, i.e. $R=0$. As for the Ricci square $R^2$ and the Kretschmann scalar, they, respectively, take the forms 
\begin{eqnarray}\label{eq:R2}
    R^{2}=
\frac{32 \pi ^2 r_s^8 \rho_s ^2 \left[(\gamma ^2-4 \gamma +8)r_s^2 +4 \gamma r_s   r+8 r^2\right]}{\big(\frac{r}{r_s+r}\big)^{2\gamma}(r_s+r)^{10}}\, ,
\end{eqnarray}
and  
\begin{eqnarray}\label{eq:K}
\begin{aligned}
K\;&= \frac{48 M^2}{r^6}
 - \frac{64 \pi M r_s^3 }{(\gamma - 3)\, r^{8}}\left(\frac{r_s + r}{r}\right)^{\gamma - 5}\times\\
&\quad
 \Bigl[
   6 r^2
   + 6(\gamma - 1) r r_s+(\gamma - 1)\gamma r_s^2
 \Bigr]\rho_s  + \mathcal{O}(\rho_s^2)\, ,
\end{aligned}
\end{eqnarray}
which reduces to the Schwarzschild case, i.e., $K=\frac{48M^2}{r^6}$, in the absence of DM halo. From these expressions, we infer that the spacetime singularity is present at $r=0$. The radial behavior of the curvature invariants is illustrated in Fig.~\ref{Fig.Curvature}, where the singular nature of the spacetime is clearly evident. Specifically, the curvature invariants vanish asymptotically as $r\to \infty$, while they diverge as $r\to 0$
\begin{eqnarray}
   \lim\limits_{r\to \infty}R,
     \vspace{3mm}R_{\mu\nu}R^{\mu\nu},
     \vspace{3mm} R_{\mu\nu\alpha\beta}R^{\mu\nu\alpha\beta}\to0\, ,
    \end{eqnarray}
and 
\begin{eqnarray}
 \lim\limits_{r\to 0} \,
    R,\, R_{\mu\nu}R^{\mu\nu}, \,
R_{\mu\nu\alpha\beta}R^{\mu\nu\alpha\beta}\to\infty\, .
\end{eqnarray}
%%%%
\begin{figure*}[ht!]

\includegraphics[width=0.245\textwidth]{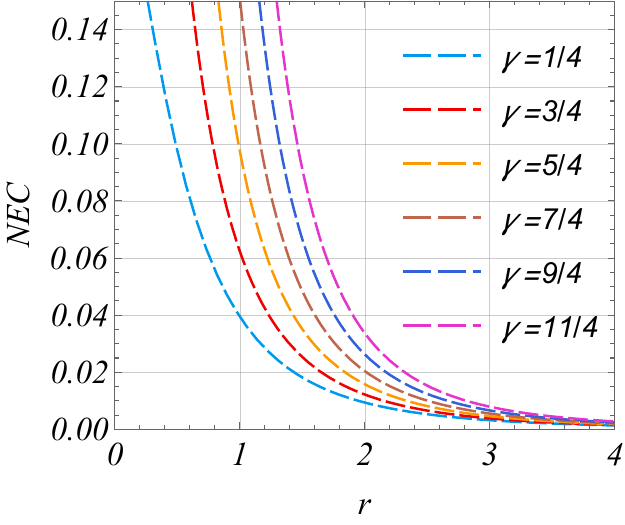}
\includegraphics[width=0.245\textwidth]{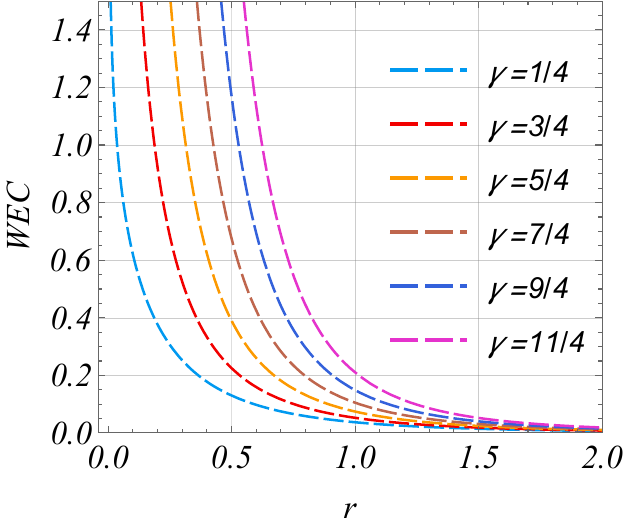}
\includegraphics[width=0.245\textwidth]{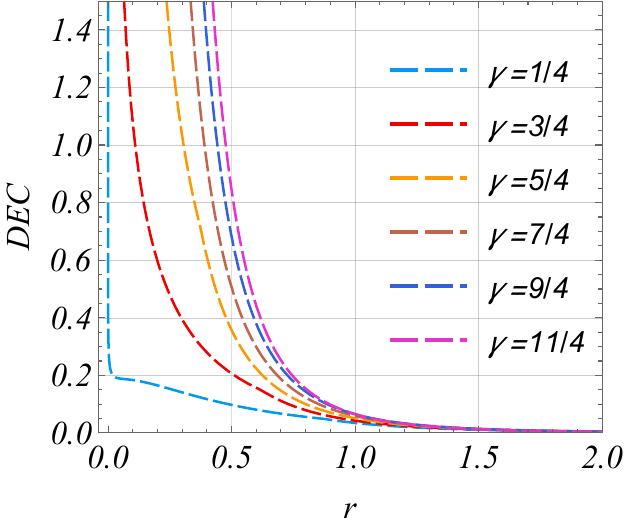}
\includegraphics[width=0.245\textwidth]{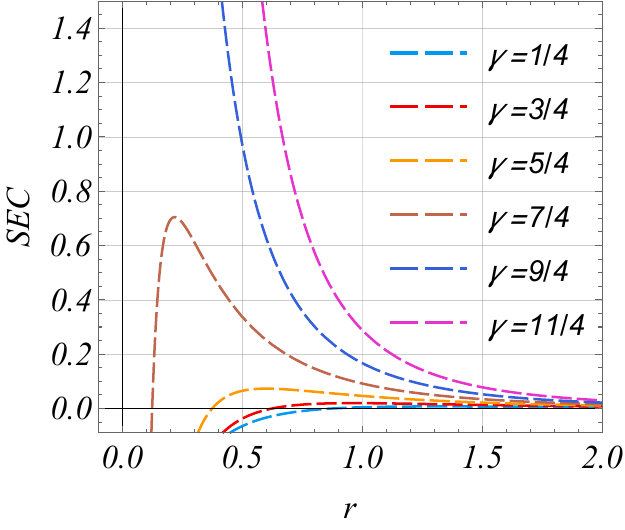}
\caption{The radial profiles of the NEC, WEC, DEC and SEC for $r_s=1$ and $\rho_s=0.5$, with different values of $\gamma$ (from left to right). 
\label{Fig.EC}}
\end{figure*}

Similarly to the curvature invariants, the energy conditions represent important fundamental properties of spacetime. Here, we examine the relevant energy conditions to gain further insight into the physical nature of the BH-DM system. Subsequently, we proceed to solve Einstein’s equations Eq.~\eqref{eq.Einstein 1} in conjunction with the corresponding energy–momentum tensor $T_{\mu}^\nu=\text{diag}[-\rho(r),P_r(r),P_{\theta}(r),P_{\phi}(r)]$. Consequently, we derive the corresponding stress-energy tensor elements as
\begin{eqnarray}\label{eq.elements of the energy-mom.}
    \rho(r)=-P_r(r)=\left(\frac{r_s}{r}\right)^4\left(\frac{r+rs}{r}\right)^{\gamma-4}\rho_s\, ,
\end{eqnarray}
\begin{eqnarray}\nonumber
    P_{\theta}(r)=P_\phi(r)=\frac{r_s^4  \left((\gamma -2) r_s+2 r\right)\rho _s}{2 \left(r_s+r\right){}^5 \left(\frac{r}{r_s+r}\right){}^{\gamma }}\, .
\end{eqnarray}
Based on the above equations, we further analyze the energy conditions associated with the DM halo source fluid. It is well known that the stress–energy tensor must satisfy a set of standard inequality constraints. Accordingly, we examine the null, weak, dominant, and strong energy conditions in this context:
\begin{itemize}
\item The null energy condition (NEC) is a fundamental requirement in general relativity, constraining the energy–momentum tensor such that the energy density along null directions remains non-negative, thereby excluding exotic phenomena associated with negative energy densities. Hence, the NEC leads to the condition \cite{2017EPJC...77..542T,Al-Badawi25JCAP}:
\begin{eqnarray}\label{eq.NEC}
      T_{\mu\nu}n^\mu n^\nu\geq0,
  \end{eqnarray}
  or
  \begin{eqnarray}\nonumber
      \rho(r)+P_i(r)\geq0, \,\,\,(i=r,\theta,\phi),
  \end{eqnarray}
 where $n^\alpha$ refers to the null vector. From Eq.~(\ref{eq.elements of the energy-mom.}),  $P_r(r)+\rho(r)=0$ reads
\begin{eqnarray}\label{eq.NEC1}
     \rho(r)+P_\beta(r)=\frac{r_s^4  \left(\gamma  r_s+4 r\right)\rho _s}{2 \left(r_s+r\right){}^5 \big(\frac{r}{r_s+r}\big)^\gamma }\geq0\, ,
  \end{eqnarray}
  where $\beta=\theta,\phi$. As can be seen from Eq.~(\ref{eq.NEC1}), it is obvious that it diverges as $r\to 0$, while it vanishes asymptotically as $r\to \infty$   
  \begin{eqnarray}\label{eq.NEC2}\nonumber
      \lim_{r\to0}[\rho(r)+P_{\beta}(r)]=\infty,
  \end{eqnarray}
  \begin{eqnarray}
      \lim_{r\to\infty}[\rho(r)+P_{\beta}(r)]=0\, .
  \end{eqnarray}
  The behavior of the NEC is demonstrated in Fig~\ref{Fig.EC} for various specific values of parameter $\gamma$ while keeping DM halo parameters $r_s$ and $\rho_s$ fixed.
  \item Similarly, the weak energy condition (WEC) is a fundamental constraint in general relativity, requiring that the energy density measured by any timelike observer remains non-negative, i.e.,  
\begin{eqnarray}\label{eq.WEC}
\rho\geq0,\,\,\rho(r)+P_i(r)\geq0, \,\,\,(i=r,\theta,\phi).
  \end{eqnarray}
 We show the behavior of the WEC in Fig.~\ref{Fig.EC} for various values of $\gamma$ for the fixed $r_s$ and $\rho_s$. 
  \item Next, we consider the dominant energy condition (DEC). This condition requires that for any future-directed causal vector field $\mathbf{Y}^\nu$, the vector $- T^\mu_\nu \mathbf{Y}^\nu$ is likewise future-directed causal vector. As a stronger condition than the WEC, the DEC physically implies that the local energy density is non-negative and that energy flux propagation is restricted to subluminal speeds for all local observers. Specifically, the DEC is satisfied provided that  
  \begin{eqnarray}\nonumber
      \rho(r)-|P_{\theta,\phi}|\geq0,
    \end{eqnarray}
  Together with Eq.~\eqref{eq.elements of the energy-mom.}, this condition can be defined by  
    \begin{eqnarray}\label{Eq.DEC}
      \frac{r_s^4 \rho _s \left(\frac{r_s+r}{r}\right){}^{\gamma }}{r_s+r}\left(r+r_s-\Big| r+\frac{1}{2} (\gamma -2) r_s\right| \Big)\geq0\, .
  \end{eqnarray}
Similar behavior compared to previous energy conditions is also expected for various values of $\gamma$ for the fixed $r_s$ and $\rho_s$, as shown in Fig.~\ref{Fig.EC}.  
  \item Finally, we consider the strong energy condition (SEC), which is determined by
\begin{eqnarray}\label{eq.SEC}
      \left(T_{\mu\nu}-\frac{1}{2}Tg_{\mu\nu}\right)n^\mu n^\nu\geq0\, \, \mbox{~~or~~}\,\,
 \rho+\sum_{i=1}^3P_i\geq0\, ,
  \end{eqnarray}
  which gives 
\begin{eqnarray}\label{eq.SEC1}
      \frac{r_s^4  \left((\gamma -2) r_s+2 r\right)\rho _s}{\left(r_s+r\right){}^5 \left(\frac{r}{r_s+r}\right){}^{\gamma }}\geq0\, .
  \end{eqnarray}
\end{itemize}

 \cred{
 We analyze the energy conditions analytically for all $r>0$, while Fig.~\ref{Fig.EC} illustrates their behavior over a representative radial range. The NEC, WEC, and DEC are satisfied for all considered values of $\gamma$. For the SEC, Eq.~\eqref{eq.SEC1} shows that it is satisfied everywhere for $\gamma\geq2$. For $\gamma<2$, the SEC is violated only in the finite region
\begin{equation*}
    0<r<r_{\rm SEC}, \qquad r_{\rm SEC}=\frac{2-\gamma}{2}r_s,
\end{equation*}
and is satisfied for $r\geq r_{\rm SEC}$. Therefore, in the exterior spacetime $r\geq r_h$, the SEC is violated only in the region $r_h<r<r_{\rm SEC}$ when $r_{\rm SEC}>r_h$. If $r_{\rm SEC}\leq r_h$, the violation lies entirely inside the event horizon.
 }

\begin{figure*}[ht!]
\includegraphics[width=0.47\textwidth]{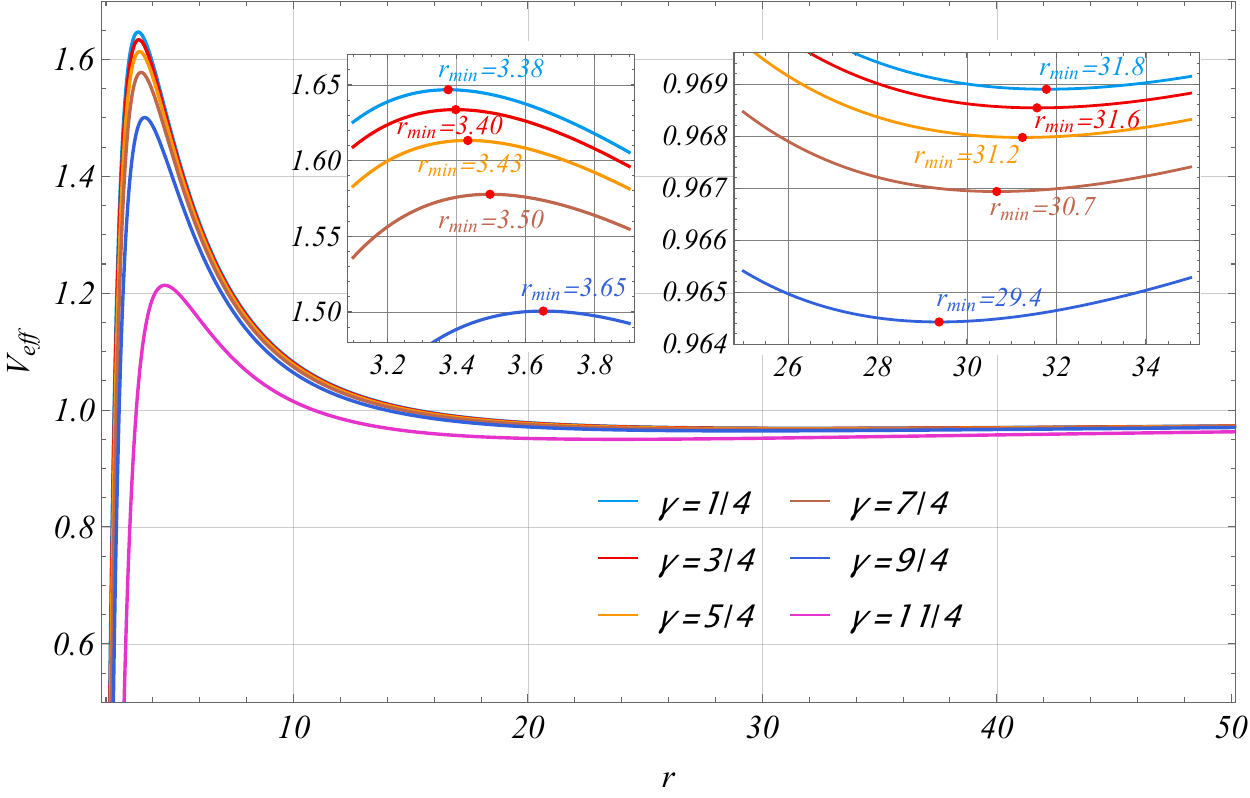}
\includegraphics[width=0.47\textwidth]{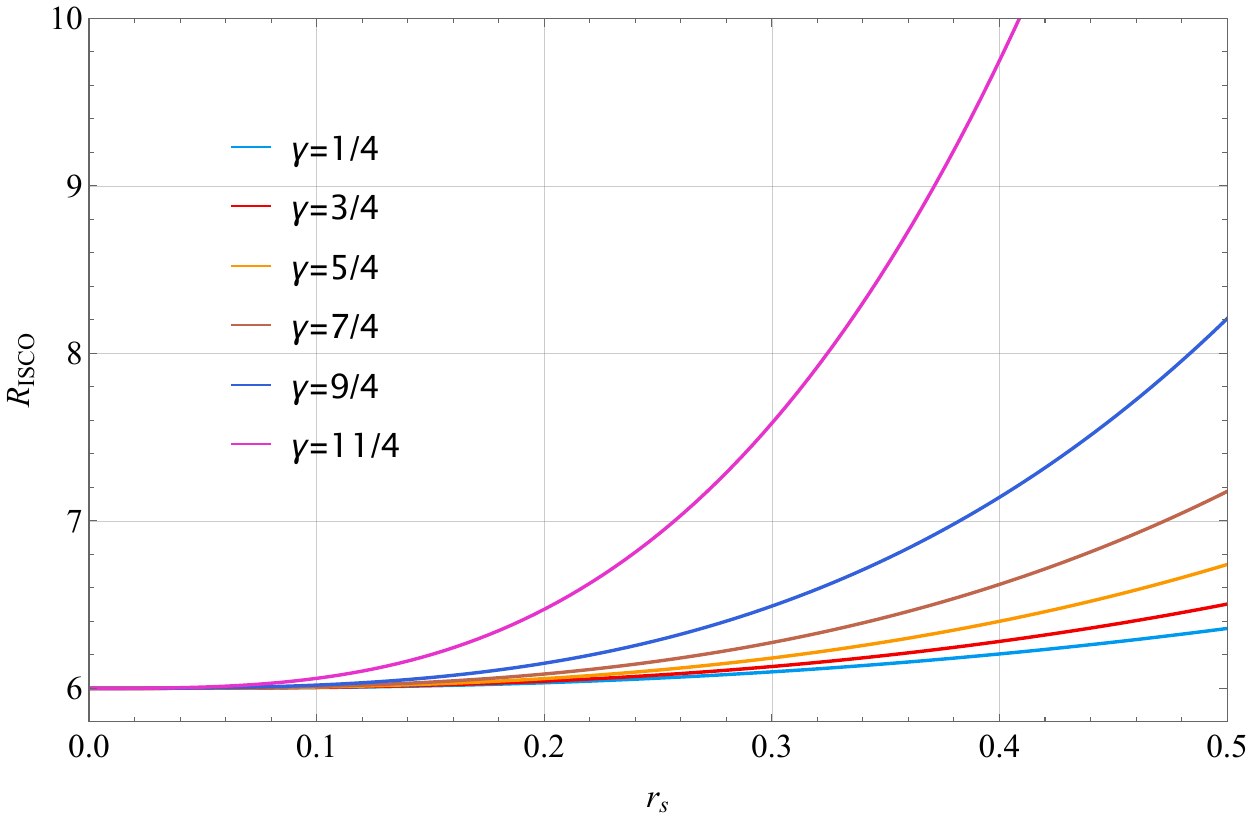}
\caption{The effective potential $V_{\rm eff}$ for the fixed $r_s$ and $\rho_s$ (left) and $R_{\rm ISCO}$ as a function of $r_s$ for the fixed $\rho_s$ (right) for various combinations of $\gamma$. 
\label{Fig.Veff}}
\end{figure*}
\begin{figure}
    \centering
    \includegraphics[width=0.6\linewidth]{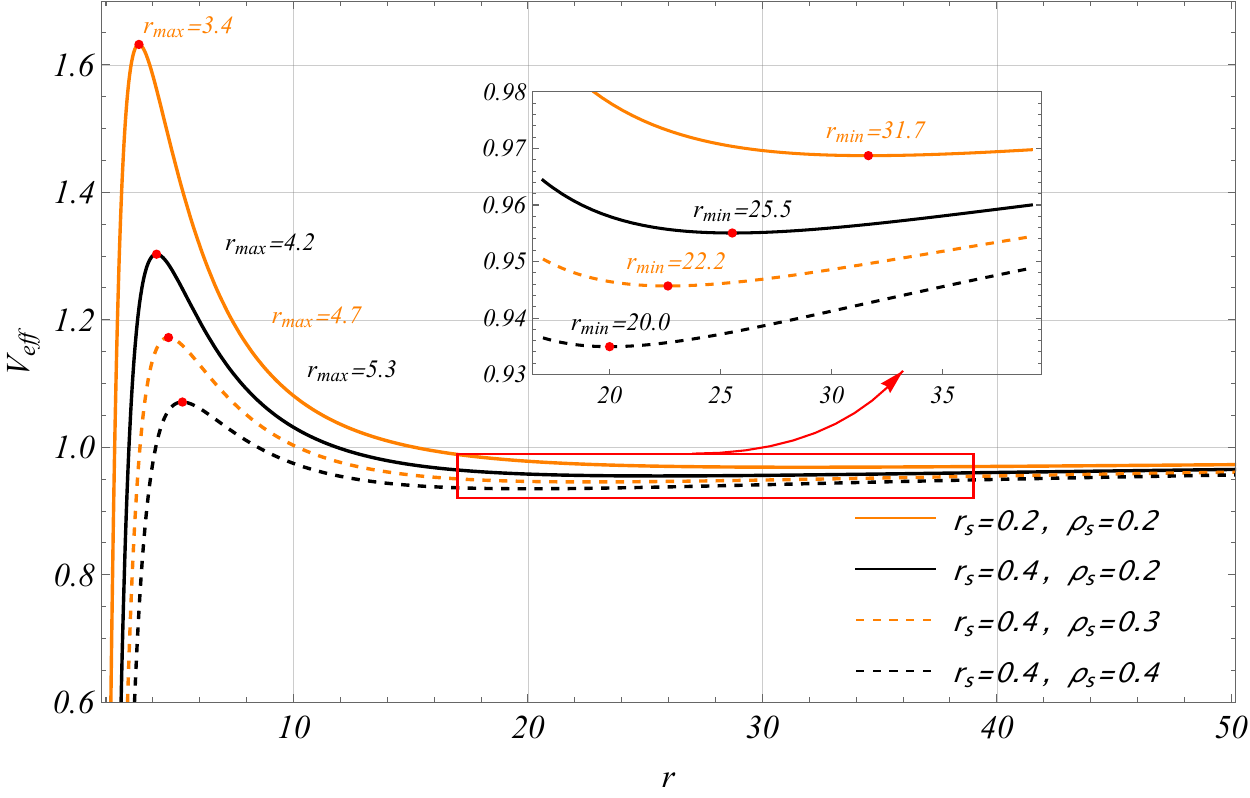
    }
    \caption{Effective potential $V_{\rm eff}(r)$ for a massive particle for various combinations of $r_s$ and $\rho_s$.  Note that $(\alpha,\beta,\gamma)=(1,4,9/4)$ is considered for the density profile. }
    \label{Fig.Veff2}
\end{figure}

%%%%%%%%%%%%%%%%%%%%%%%%%%%%%%%%%%%%%%%%%%%%%%%%%%%%%%%%%%
\section{Timelike and null geodesics }\label{Sec:IV}

To explore the impact of a DM halo on spacetime, we analyze the motion of massive particles and photons around a Schwarzschild-like BH embedded in a DM halo. Furthermore, we constrain the halo parameters $r_s$ and $\rho_s$ using observational data from the perihelion shifts of Mercury and the S2 star, together with the shadow measurements of M87$^{\ast}$ and Sgr A$^{\ast}$.
The dynamics of a particle in the spacetime of a Schwarzschild-like BH described by the metric \eqref{eq.full-line} can be formulated within the Hamiltonian formalism \cite{Misner73}. The Hamiltonian is given by
\begin{equation}\label{hamiltonian}
H=\frac{1}{2}g^{\mu\nu}p_\mu p_\nu \ ,
\end{equation}
where $p^\mu=m\,u^\mu$ is the four-momentum and $u^\mu = dx^\mu/d\tau$ represents the four-velocity of the particle. In spherical coordinates, the indices $\mu$ and $\nu$ run over the coordinate components $(t, r, \theta, \phi)$. For massive particles $H=-m^2/2$, while for photons $H=0$. The equations of motion are obtained from Hamilton's equations
\begin{equation} \label{Hamx}
    \frac{dx^\mu}{d\lambda} = \frac{\partial H}{\partial p_\mu} \quad \text{and}\quad \frac{dp_\mu}{d\lambda} = -\frac{\partial H}{\partial x^\mu} \ ,
\end{equation}
where $\lambda=\tau/m$ is the affine parameter and $\tau$ represents the proper time of the massive particle.

The spacetime metric \eqref{eq.full-line} is independent of the coordinates $t$ and $\phi$, leading to the conservation of the particle’s energy and angular momentum, given by $p_t=-E$ and $p_\phi=L$ \cite{Misner73}.

\subsection{Dynamics of massive particles}

\begin{figure*}
    \centering
    \includegraphics[scale=0.45]{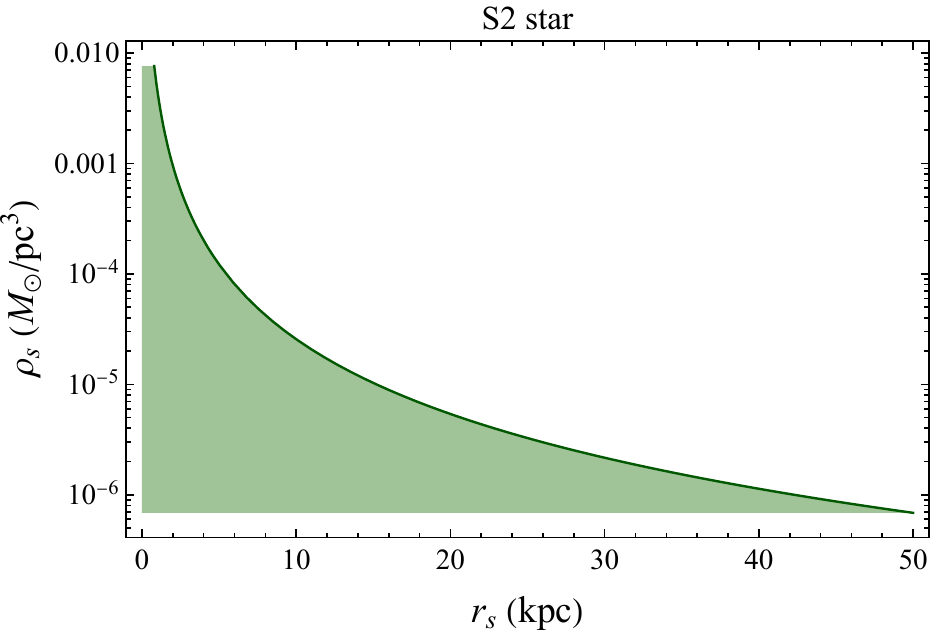}
    \includegraphics[scale=0.45]{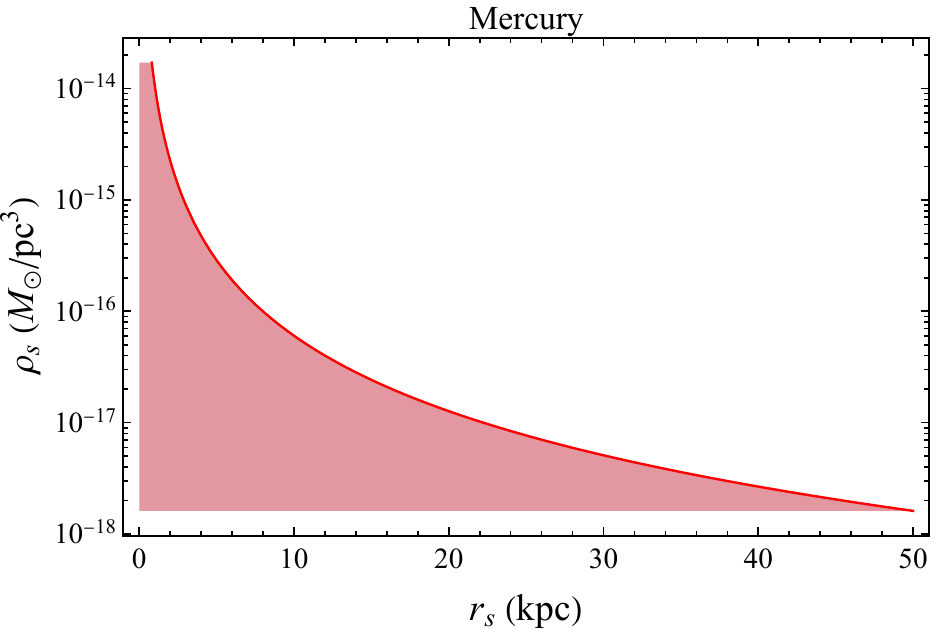}
    \caption{
    Parameter space of the halo radius $r_s$ and characteristic density $\rho_s$. The shaded regions represent the observationally allowed values of $(r_s,\rho_s)$ inferred from S2 star (left panel) and Mercury (right panel) measurements.
 }
    \label{fig:parspace}
\end{figure*}

Using Eq.~\eqref{hamiltonian} and conservation laws, the Hamiltonian for a massive particle can be written as
\begin{align} \label{ex-hamiltonian}
H = \frac{1}{2} \left( g^{rr} p_r^2 + g^{\theta\theta} p_\theta^2 + g^{tt} E^2 + g^{\phi\phi} L^2 \right) = -\frac{m^2}{2}.
\end{align}
For simplicity, we assume that the massive particle moves in the equatorial plane $\theta=\pi/2$. From Eq.~\eqref{eq.full-line}, $g^{tt}=-1/f(r)$, $g^{rr}=f(r)$, and $g^{\phi\phi}=1/r^2$. Substituting into Eq.~\eqref{ex-hamiltonian}, we obtain the equations governing the radial and azimuthal motion:
\begin{align} \label{radial}
   \left(\frac{dr}{d\tau}\right)^2 
   = \mathcal{E}^2-f(r)\left(1+\frac{\mathcal{L}^2}{r^2}\right) \quad \text{and}\quad \frac{d\phi}{d\tau} = \frac{\mathcal{L}}{r^2} \, ,
\end{align}
where $\mathcal{E}=E/m$ and $\mathcal{L}=L/m$ represent the specific energy and specific angular momentum, respectively. From Eq.~\eqref{radial}, the effective potential takes the general form
\begin{eqnarray}
    V_{\rm eff}(r) &=& \left[1-\frac{2M}{r}-\frac{8\pi\rho_sr_s^3}{(3-\gamma)r}\Big(\frac{r}{r+r_s}\Big)^{3-\gamma}\right]\times\nonumber\\
    &&\left(1+\frac{\mathcal{L}^2}{r^2}\right)\, .
\end{eqnarray}
Here, we choose the following density profile $(\alpha,\beta,\gamma)=(1,4,9/4)$, for which the effective potential becomes
\begin{eqnarray}\label{Veff_1494}
    V_{\rm eff}^{^{(1,4,9/4)}}(r) &=&\left[ 1-\frac{2M}{r}-\frac{32\pi\rho_s r_s^3}{3r}\left(\frac{r}{r+r_s}\right)^{3/4}\right]\times\nonumber\\
    &\times&\left(1+\frac{\mathcal{L}^2}{r^2}\right)\, .
\end{eqnarray}
Using the above effective potential, the ISCO radius can be obtained from the conditions $\frac{dV_{\rm eff}}{dr}=0$ and $\frac{d^2V_{\rm eff}}{dr^2}=0$ \cite{Dadhich22a,Dadhich22IJMPD}.

Fig.~\ref{Fig.Veff} shows the effective potential and $R_{\rm ISCO}$ for the density profile $(\alpha,\beta,\gamma)=(1,4,\gamma)$. 
As seen from the figure, with increasing $\gamma$, the height of the effective potential decreases, and the location of the unstable orbit (corresponding to the maximum of the effective potential) shifts toward larger values of $r$. 
In addition, $R_{\rm ISCO}$ increases with increasing $\gamma$ for a given value of $r_s$. This suggests that the gravitational effects become stronger as $\gamma$ increases. Fig.~\ref{Fig.Veff2} shows the effective potential for the density profile $(1,4,9/4)$ for different values of DM halo parameters, i.e., $r_s$ and $\rho_s$. We observe that with increasing $r_s$ and $\rho_s$, the unstable orbits shift outward, while the stable orbits (minima of the effective potential) shift inward.

By combining Eqs.~\eqref{radial} and \eqref{Veff_1494}, the trajectory equation can be expressed as
\begin{eqnarray} \label{trajectory-eqn}
    \left(\frac{dr}{d\phi}\right)^2 
    = \frac{r^4\left(  \mathcal{E}^2 
      - V_{\rm eff}^{^{(1,4,9/4)}}(r)\right)}{\mathcal{L}^2}  \, .
\end{eqnarray}
Introducing the transformation $u=1/r$ and differentiating with respect to $\phi$, the trajectory equation becomes
\begin{eqnarray}\label{geod-eq}
    \frac{d^2 u}{d\phi^2} =\frac{M}{\mathcal{L}^2} - u + \frac{g(u)}{\mathcal{L}^2} \, ,
\end{eqnarray}
where 
\begin{eqnarray}
    \frac{g(u)}{\mathcal{L}^2} &=& 3 M u^2+4 \pi  \rho_s r_s^3 \left(\frac{1}{1+r_s u}\right)^{7/4}\times\nonumber\\
    &\times&\frac{ \left(4+u \left(r_s+12 u L^2+9 r_s u^2 L^2 \right)\right)}{3 L^2} \, .
\end{eqnarray}
Adopting the procedure outlined in Ref.~\cite{Adkins_2007}, the perihelion shift after one orbital period is given by
\begin{equation} \label{p-shift-eqn}
    \Delta \varphi = \frac{\pi}{\mathcal{L}^2} \left| \frac{d g(u)}{du} \right|_{u = \frac{1}{b}}\, ,
\end{equation}
where $b = a(1-e^2)$, in which $a$ and $e$ correspond to the semi-major axis and the eccentricity of the orbit. 

\cred{
In order to express the halo parameters in physical units, we introduce the corresponding dimensionless quantities entering the perihelion-shift formula. The required transformations are
\begin{align}
M_{\rm g} &= \frac{GM}{c^2}, \\
\mathcal{L}^2 &= M_{\rm g}\,a(1-e^2), \\
\hat{\rho}_s &=
\frac{32\pi G^3 M^2 M_\odot}
{3c^6\,{\rm pc}^3}
\left[
\frac{\rho_s}{M_\odot\,{\rm pc}^{-3}}
\right], \\
\hat{r}_s &=
\frac{c^2\,{\rm kpc}}{GM}
\left[
\frac{r_s}{{\rm kpc}}
\right].
\end{align}
where the quantities in square brackets denote the numerical values of $\rho_s$ and $r_s$ expressed in $M_\odot\,{\rm pc}^{-3}$ and kpc, respectively, while $\hat{\rho}_s$ and $\hat{r}_s$ are the corresponding dimensionless quantities.
}

Using the above equations together with Eq.~\eqref{p-shift-eqn}, the perihelion shift after one complete revolution can be written as

\begin{align} \label{eq:perishift}
\Delta\phi_{\rm th}
&=6\pi\alpha
+\frac{3\pi\alpha}{32}\,
\hat{\rho}_s\hat{r}_s^3
\left(\frac{1}{1+\alpha\hat{r}_s}\right)^{11/4}
\nonumber\\
&\quad\times
\left[
32-8(1-5\alpha)\hat{r}_s
-\alpha(1-15\alpha)\hat{r}_s^2
\right],
\end{align}
where
\begin{equation}
\alpha = \frac{G M}{a c^2 (1 - e^2)}.
\end{equation}

\textcolor{black}{
We now consider an astronomical application of the perihelion shift. In particular, we use the orbital data of Mercury and the S2 star to examine the allowed parameter space of $\rho_s$ and $r_s$. For the S2 star, this analysis provides a constraint on the DM distribution surrounding Sgr A*. In the case of Mercury, however, we do not assume the existence of a realistic Dehnen-type DM halo centered on the Sun. Instead, the precisely measured perihelion advance of Mercury is used as a phenomenological weak-field test to constrain possible deviations from the Schwarzschild geometry predicted by the present model. The orbital parameters and observed perihelion shifts adopted in our analysis are given below.
}

\vspace{1em}
\noindent\textbf{\textit{Mercury:}}
\begin{eqnarray*}
    \frac{2 G M_\odot}{c^2} &=& 2.95325008 \times 10^3 \, [\text{m}] \, , \nonumber \\
    a &=&  5.7909175 \times 10^{10} \, [\text{m}] \, , \nonumber \\
    e &=&  0.20563069\, , \nonumber \\
    \Delta \phi_{\text{obs}} &=& 2\pi \times (7.98734 \pm 0.00037) \times 10^{-8}~\mathrm{rad/rev}.
\end{eqnarray*}
\vspace{1em}
\noindent\textbf{\textit{S2 star:}}
\begin{eqnarray*}
    M_{\text{Sgr A}^*} &=& 4.260 \times 10^6 M_{\odot}\, , \\
    a_{\text{S2}} &=& 970 \, [\text{au}] \, , \\
    1 \ \text{au} &=& 1.495978707\times10^{11} \, [\text{m}]\, , \\
    e_{\text{S2}} &=& 0.884649 \, , \\
    T_{\text{S2}} &=& 16.052 \, [\text{years}] \, , \\
    \Delta\phi_{\text{obs}} &=& 48.298 \ f_{\text{SP}} \ \Big[\ ^{\prime\prime}/\text{year} \Big] \, ,\, f_{\text{SP}} = 1.10 \pm 0.19  \, .
\end{eqnarray*}

\textcolor{black}{
Here, the reported uncertainties are the $1\sigma$ statistical uncertainties of the corresponding observations. For each point in the $(\rho_s,r_s)$ plane, we calculate the theoretical perihelion shift $\Delta\phi_{\rm th}$ from Eq.~\eqref{eq:perishift} using the orbital parameters of Mercury or the S2 star, respectively. The theoretical prediction is considered consistent with the observation when it lies within the $1\sigma$ observational interval,
\begin{eqnarray}
\left|\Delta\phi_{\rm th}-\Delta\phi_{\rm obs}\right|
\leq \sigma_{\Delta\phi},
\end{eqnarray}
where $\sigma_{\Delta\phi}$ denotes the $1\sigma$ uncertainty of the observed perihelion shift. 
}

\textcolor{black}{
Using this criterion together with Eq.~\eqref{eq:perishift}, we obtain the allowed regions in the $(\rho_s,r_s)$ parameter space shown in Fig.~\ref{fig:parspace}. For fixed values of $r_s=5$, $10$, and $20~{\rm kpc}$, we obtain
the corresponding $1\sigma$ upper bounds on $\rho_s$ from the
perihelion-shift measurements. These values are listed in
Table~\ref{tab:perihelion_bounds}. Both Fig.~\ref{fig:parspace} and Table~\ref{tab:perihelion_bounds} show that the S2-star analysis allows larger values of $\rho_s$ than the Mercury analysis for the same values of $r_s$. This shows that the effect of the DM halo is more pronounced around more massive objects.}

\renewcommand{\arraystretch}{1.5}
\begin{table}[ht]
\centering
\begin{tabular}{ccc}
\hline\hline
$r_s$ (kpc) &
$\rho_{s,\max}^{\rm Mercury}$ ($M_\odot\,{\rm pc}^{-3}$) &
$\rho_{s,\max}^{\rm S2}$ ($M_\odot\,{\rm pc}^{-3}$) \\
\hline
5  & $2.87\times10^{-16}$ & $1.22\times10^{-4}$ \\
10 & $6.03\times10^{-17}$ & $2.56\times10^{-5}$ \\
20 & $1.27\times10^{-17}$ & $5.39\times10^{-6}$ \\
\hline\hline
\end{tabular}
\caption{Representative $1\sigma$ upper bounds on the characteristic
density $\rho_s$ obtained from the perihelion-shift analysis for fixed
values of the scale radius $r_s$.}
\label{tab:perihelion_bounds}
\end{table}

\begin{figure*}
    \centering
\includegraphics[scale=0.4]{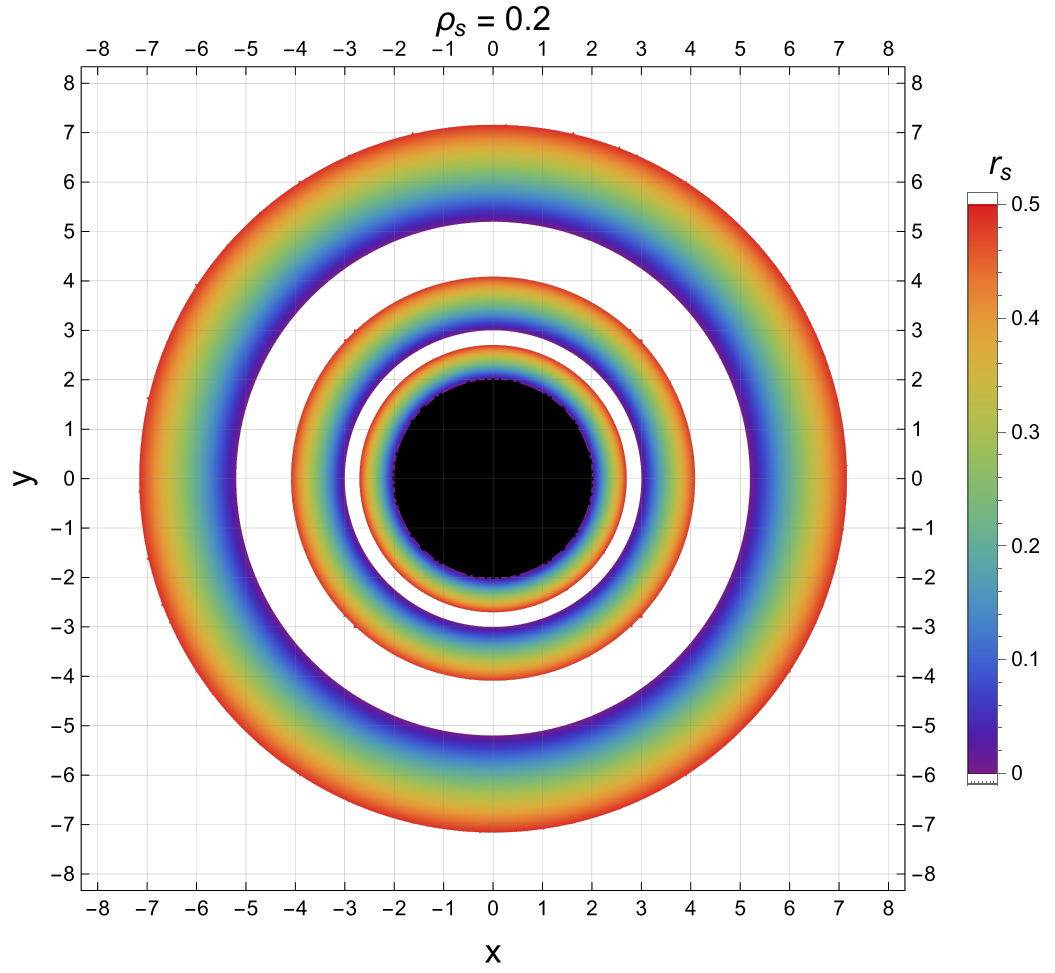}
\includegraphics[scale=0.4]{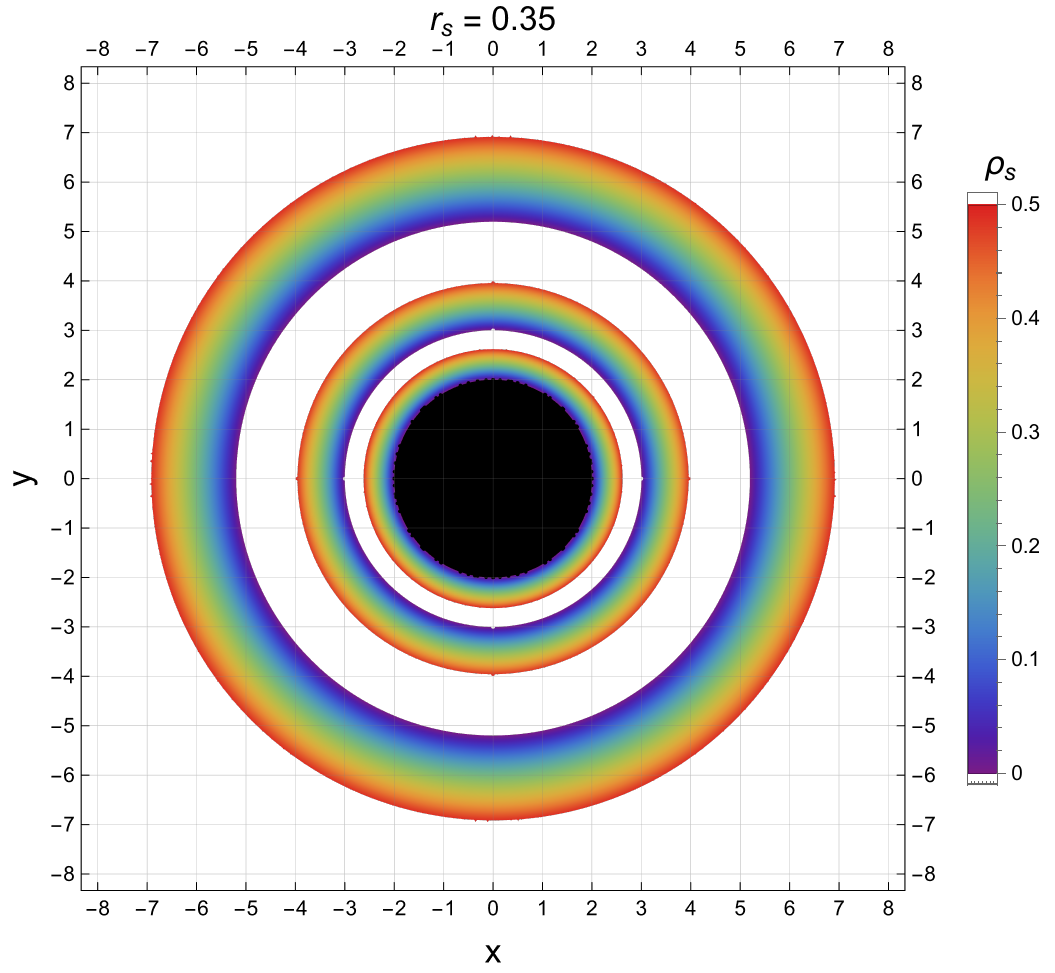}    
    \caption{Shadow structure of a Schwarzschild-like BH in the presence of a DM halo. The central black disk represents the event horizon of the Schwarzschild BH. The innermost colored ring illustrates how the horizon radius varies with the DM halo parameters, while the middle ring corresponds to the photon sphere. The outermost ring represents the BH shadow. The color scale indicates the values of the DM halo parameters: the scale radius $r_s$ (left panel) and the characteristic density $\rho_s$ (right panel). }
    \label{Fig.shadow}
\end{figure*}

\subsection{Photon geodesics}

We now investigate the null geodesics around the Schwarzschild-like BH surrounded by DM halo characterized by $(\alpha,\beta,\gamma)=(1,4,9/4)$. The four-momentum of a photon can be written as
\begin{equation} \label{momentum-photon}
    p^\alpha = dx^\alpha/d\lambda \, .
\end{equation}  

As in the massive particle case, we restrict the motion to the plane $\theta = \pi/2$. Using Eqs.~\eqref{hamiltonian}, \eqref{Hamx} together with the conservation laws, we obtain the following equation governing the propagation of light.
\begin{eqnarray}
    \frac{dt}{d\lambda}=\frac{E}{f(r)} \mbox{~~~and~~~}
    \frac{d\phi}{d\lambda}=\frac{L}{r^2},
\end{eqnarray}    
\begin{equation}
    \left(\frac{dr}{d\lambda}\right)^2=E^2-f(r)\frac{L^2}{r^2}.
\end{equation}

Introducing the transformation $\lambda'=L\lambda$, the above equations can be rewritten as
\begin{eqnarray}
    \frac{dt}{d\lambda'}=\frac{1}{bf(r)} \mbox{~~~and~~~}\frac{d\phi}{d\lambda'}=\frac{1}{r^2}\, ,
\end{eqnarray} 
\begin{equation}
    \left(\frac{dr}{d\lambda'}\right)^2=\frac{1}{b^2}-\frac{f(r)}{r^2}=V_{\rm eff}^{ph}(r)\, ,
    \label{radialM}
\end{equation}
where $b=L/E$ denotes the impact parameter, and $V_{\rm eff}^{ph}$ represents the effective potential for photons. The radius $r_{ph}$ of the photon sphere is defined by
\begin{equation}
    \frac{d}{dr}\left(V_{\rm eff}^{ph}(r)\right) = 0
\end{equation}
Since the solution of the above equation is complicated, we solve it numerically and present the radii of the photon sphere for different values of $\rho_s$ and $r_s$ in Table~\ref{tab:photonsphere}.
\begin{table}[h]
\centering
\begin{tabular}{c|cccccc}
\hline
$\rho_s \backslash r_s$ & 0.0 & 0.1 & 0.2 & 0.3 & 0.4 & 0.5 \\
\hline
0.0 & 3.0000 & 3.00000 & 3.00000 & 3.00000 & 3.00000 & 3.00000 \\
0.1 & 3.0000 & 3.00487 & 3.03774 & 3.12392 & 3.28715 & 3.55166 \\
0.2 & 3.0000 & 3.00973 & 3.07554 & 3.24867 & 3.57932 & 4.12205 \\
0.3 & 3.0000 & 3.01460 & 3.11340 & 3.37416 & 3.87559 & 4.70550 \\
0.4 & 3.0000 & 3.01946 & 3.15130 & 3.50033 & 4.17520 & 5.29827 \\
0.5 & 3.0000 & 3.02433 & 3.18926 & 3.62711 & 4.47759 & 5.89785 \\
\hline
\end{tabular}
\caption{Photon sphere radius $r_{ph}$ for different values of $\rho_s$ and $r_s$.}
\label{tab:photonsphere}
\end{table}

For a distant observer, the radius of the BH shadow is determined by the critical impact parameter of photon trajectories, $R_{sh} = b_c$. A photon coming from infinity with $b=b_c$ asymptotically approaches the photon sphere, and therefore the apparent radius of the shadow corresponds to this critical impact parameter. This critical impact parameter $b_c$ is obtained from the condition 
$V_{\rm eff}^{ph}(r_{ph}) = 0$ and is given by
\begin{equation}
R_{sh} = b_c = \frac{r_{ph}}{\sqrt{f(r_{ph})}}\, .
\end{equation}

Fig.~\ref{Fig.shadow} illustrates the shadow, photon sphere, and horizon structure of the Schwarzschild-like BH in the presence of DM halo. Table~\ref{tab:photonsphere} lists the corresponding numerical values of $r_{\rm ph}$. It is observed that the radii of the shadow, photon sphere, and horizon increase with increasing $r_s$ and $\rho_s$. This behavior indicates that the DM halo enhances the effective gravitational field around the BH.

\section{CONSTRAINTS ON BLACK HOLE AND DM HALO PARAMETERS}\label{Sec:V}

The observable quantity measured in astronomical observations is the angular diameter of the BH shadow, $\theta_{sh}$, which is related to the dimensionless shadow radius $R_{sh}$ by
\begin{equation}
\theta_{sh} = R_{sh}\times\frac{2GM}{c^2 D},
\end{equation}
where $D$ denotes the distance between the observer and the BH.

\begin{figure*}
    \centering
\includegraphics[scale=0.31]{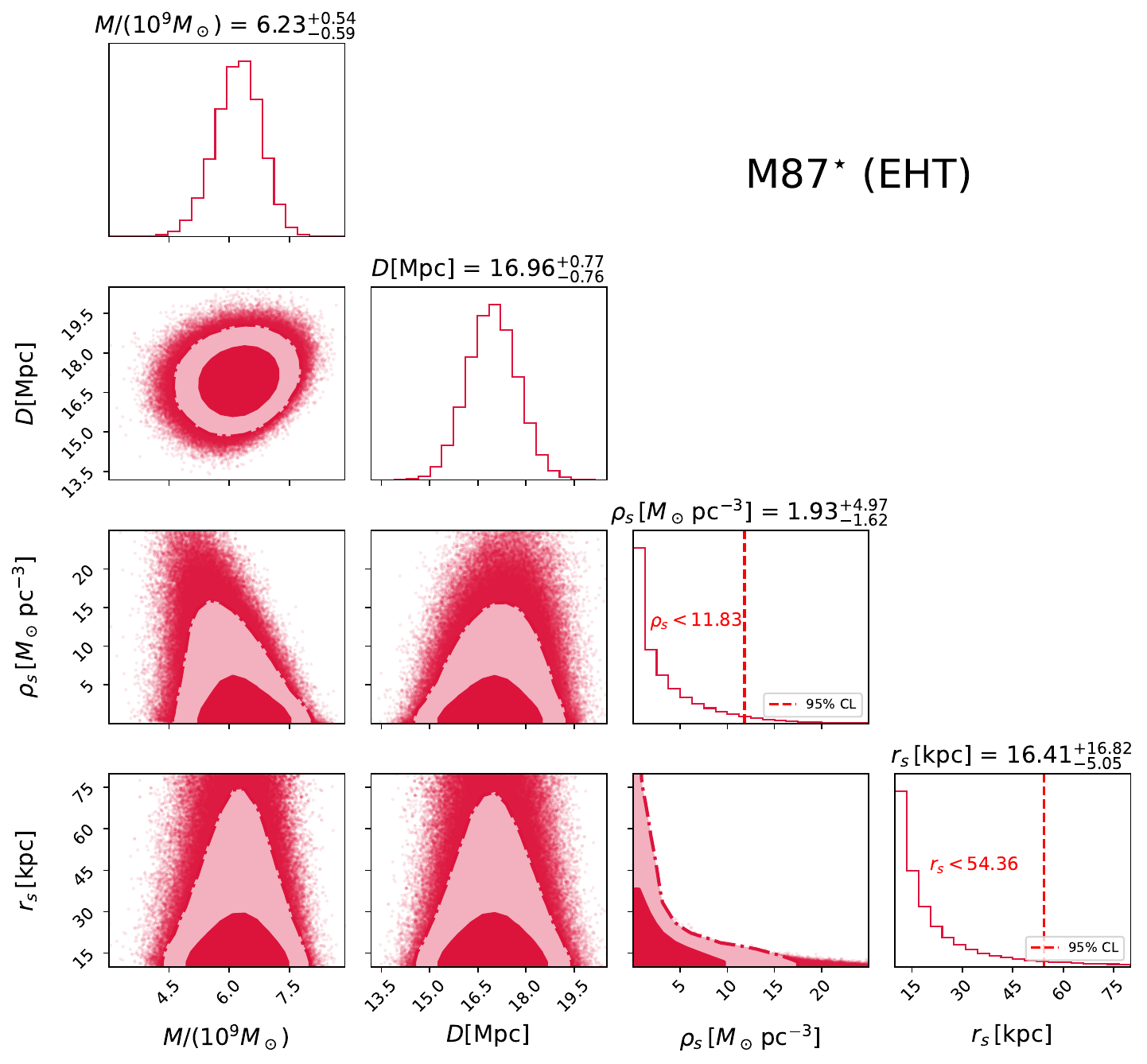}
\includegraphics[scale=0.31]{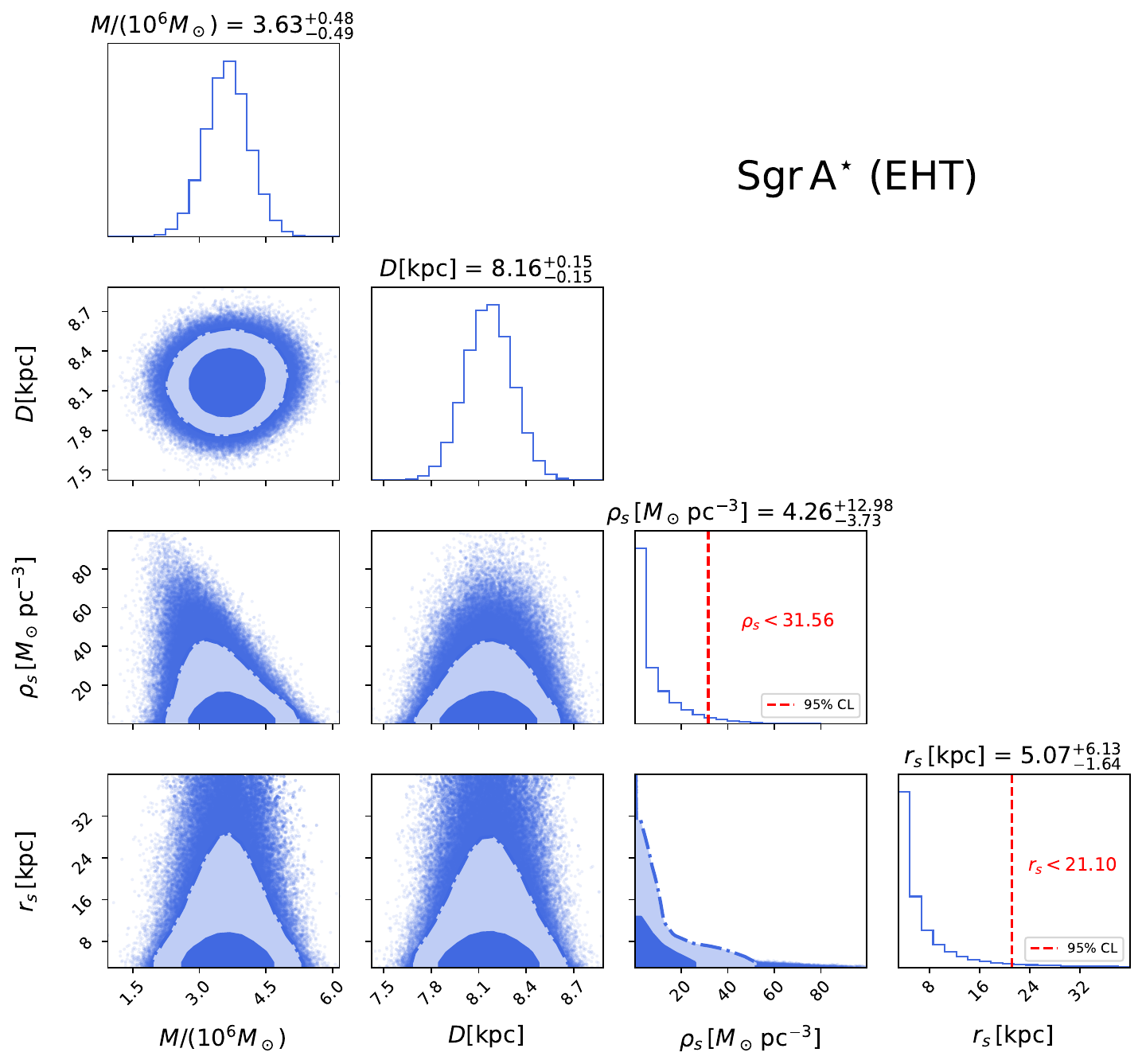}
\includegraphics[scale=0.31]{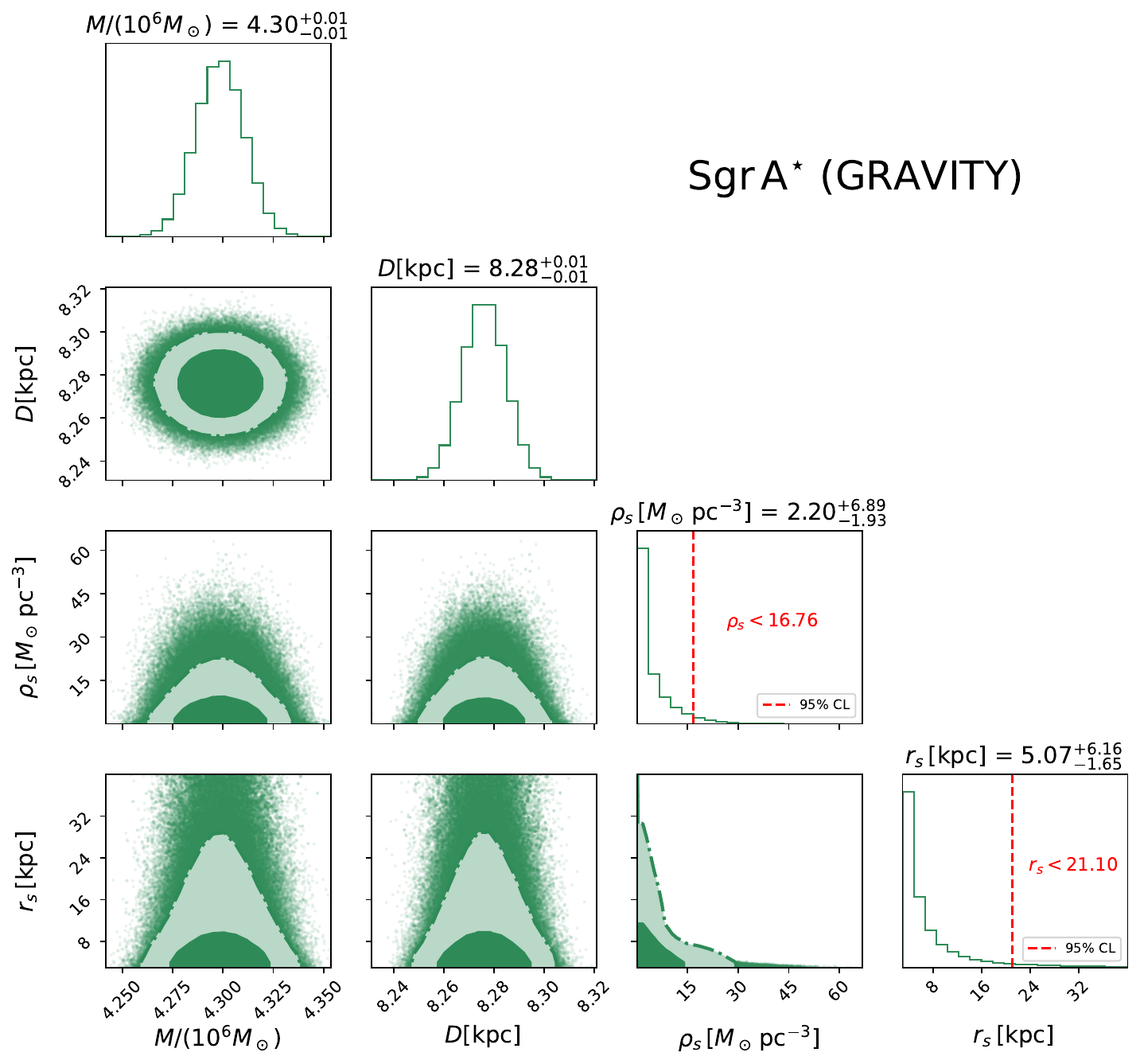}
\includegraphics[scale=0.31]{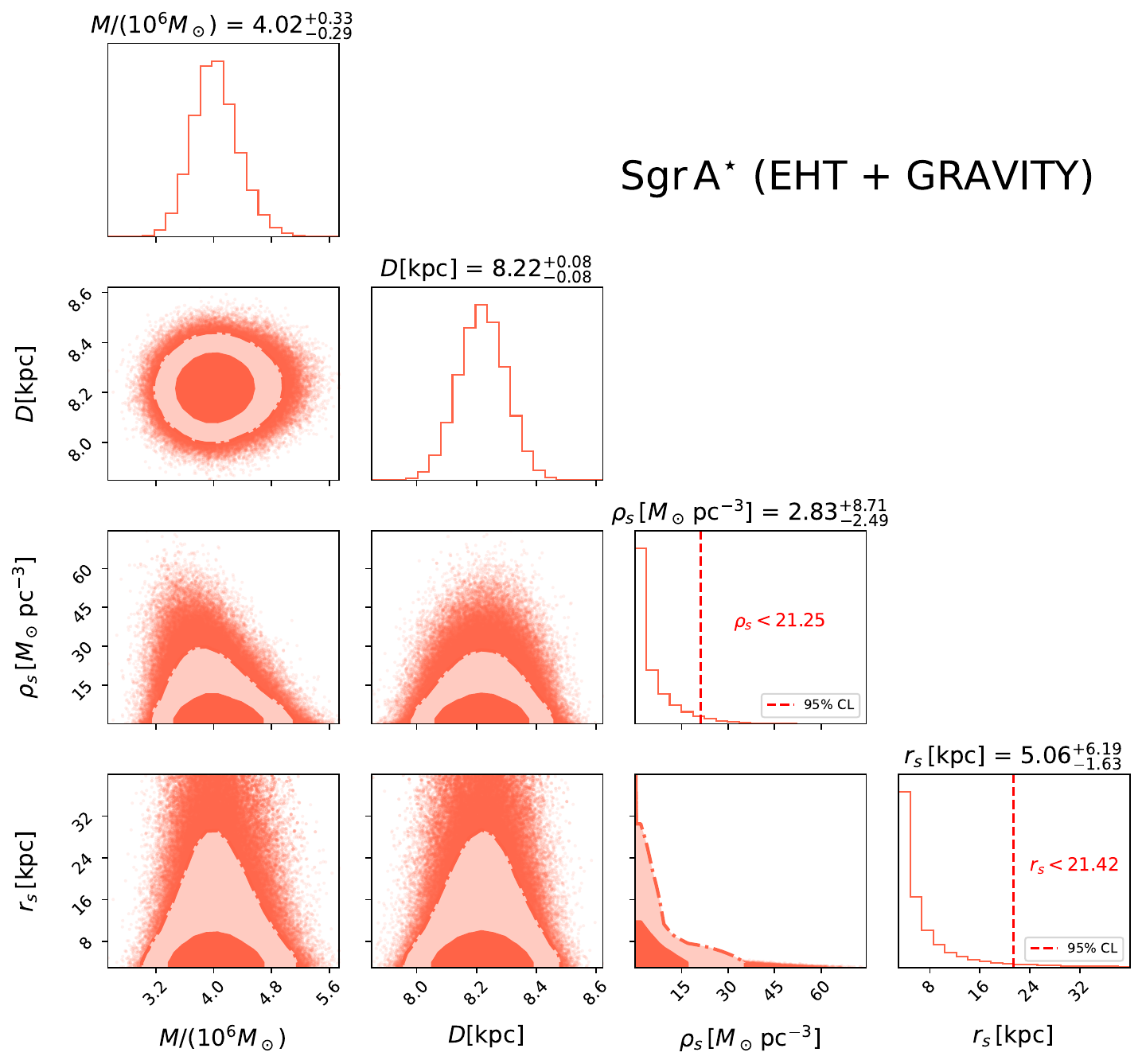}
    \caption{ Posterior probability distributions for the BH mass ($M$), distance ($D$), DM halo density ($\rho_s$), and halo scale radius ($r_s$), obtained from the shadow observations of Sgr A$^{\star}$ and M87$^{\star}$. The  red dashed vertical lines indicate the 95\% confidence levels (CL) on $\rho_s$ and $r_s$. }
    \label{Fig.mcmc}
\end{figure*}

\begin{table}[t]
\centering
\renewcommand{\arraystretch}{1.5}

\resizebox{0.8\textwidth}{!}{%
\begin{tabular}{lcccc}
\hline\hline

Object &
$\theta_{\rm sh}$ ($\mu$as) &
Distance $D$ &
Mass $M$ &
Data
\\

\hline

M87$^{\star}$
&
$42 \pm 3$
&
$16.8 \pm 0.8$ Mpc
&
$(6.5 \pm 0.7)\times10^{9}\,M_{\odot}$
&
EHT
\\

Sgr A$^{\star}$
&
$48.7 \pm 7$
&
$8150 \pm 150$ pc
&
$(4.0^{+1.1}_{-0.6})\times10^{6}\,M_{\odot}$
&
EHT
\\

Sgr A$^{\star}$
&
---
&
$8275.9 \pm 8.6$ pc
&
$(4.299 \pm 0.012)\times10^{6}\,M_{\odot}$
&
GRAVITY
\\

\hline\hline
\end{tabular}%
}

\caption{
Observed parameters for the supermassive BHs M87$^{\star}$ and
Sgr A$^{\star}$ (see details in
\cite{Akiyama22L12,Akiyama19L6,Gravity2024A&A692A242G}).
}
\label{tab:shadow}

\end{table}

\textcolor{black}{
Using the observational data for M87$^\star$ and Sgr~A$^\star$ listed in Table~\ref{tab:shadow}, we constrain the BH mass $M$, distance $D$, characteristic density $\rho_s$, and scale radius $r_s$ of the DM halo through an MCMC analysis. 
In this analysis, $M$ refers to the central BH mass defined in Sec.~\ref{Sec:II}, while the DM halo contribution is included separately through $\rho_s$ and $r_s$.
For the M87$^\star$ and Sgr~A$^\star$ EHT cases, we use the corresponding EHT measurements of the shadow angular diameter, $M$, and $D$. Since GRAVITY does not provide an independent measurement of the Sgr~A$^\star$ shadow diameter, the GRAVITY case uses the EHT shadow measurement together with the GRAVITY measurements of $M$ and $D$. For the combined EHT+GRAVITY case, we take the arithmetic mean of the EHT and GRAVITY central values of $M$ and $D$, and average the corresponding upper and lower uncertainties separately. The parameter space is explored using the Python package \textit{emcee}~\cite{Foreman_Mackey_2013}, which implements Markov Chain Monte Carlo (MCMC) sampling.
}

The likelihood function quantifies how well the theoretical prediction of the angular shadow $\theta_{\rm sh}^{\rm th}$ agrees with the observational measurement $\theta_{\rm sh}^{\rm obs}$ for a given set of model parameters. The logarithmic likelihood function is therefore written as
\begin{equation}
    \log \mathcal{L}(M,D, \rho_s, r_s) = -\frac{1}{2}  \left( \frac{\theta_{\rm sh}^{\rm th}(M, D, \rho_s, r_s) - \theta_{\rm sh}^{\rm obs}}{\sigma_{\rm sh}^{\rm obs}} \right)^2 \, ,
\end{equation}
where $\sigma_{\rm sh}^{\rm obs}$ represents the corresponding observational uncertainty.

\begin{table}[t]
\centering
\renewcommand{\arraystretch}{1.25}
\setlength{\tabcolsep}{8pt}

\resizebox{\textwidth}{!}{%
\begin{tabular}{lcccc}
\hline\hline

Dataset &
Prior on $M$ &
Prior on $D$ &
Prior on $\rho_s$ &
Prior on $r_s$
\\

\hline

M87$^\star$ EHT
&
$\mathcal{N}(6.5,0.7)\times10^9 M_\odot$
&
$\mathcal{N}(16.8,0.8)\ {\rm Mpc}$
&
$\mathcal{U}(0,25)\ M_\odot\,{\rm pc}^{-3}$
&
$\mathcal{U}(10,80)\ {\rm kpc}$
\\

Sgr A$^\star$ EHT
&
$4.0^{+1.1}_{-0.6}\times10^6 M_\odot$
&
$\mathcal{N}(8.150,0.150)\ {\rm kpc}$
&
$\mathcal{U}(0,100)\ M_\odot\,{\rm pc}^{-3}$
&
$\mathcal{U}(3,40)\ {\rm kpc}$
\\

Sgr A$^\star$ GRAVITY
&
$\mathcal{N}(4.299,0.012)\times10^6 M_\odot$
&
$\mathcal{N}(8.2759,0.0086)\ {\rm kpc}$
&
$\mathcal{U}(0,100)\ M_\odot\,{\rm pc}^{-3}$
&
$\mathcal{U}(3,40)\ {\rm kpc}$
\\

Sgr A$^\star$ EHT+GRAVITY
&
$4.1495^{+0.556}_{-0.306}\times10^6 M_\odot$
&
$\mathcal{N}(8.21295,0.0793)\ {\rm kpc}$
&
$\mathcal{U}(0,100)\ M_\odot\,{\rm pc}^{-3}$
&
$\mathcal{U}(3,40)\ {\rm kpc}$
\\

\hline\hline
\end{tabular}%
}

\caption{\textcolor{black}{
Priors adopted in the MCMC analysis of the black-hole shadow.
Gaussian priors are used for the black-hole mass $M$ and distance $D$,
while uniform priors are adopted for the dark-matter halo parameters
$\rho_s$ and $r_s$
~\cite{Oldham_2018MNRAS,ChaoLi_2020MNRAS,Xiaowei_2024MNRAS,
Cautun_2020MNRAS,Shen24PLB}.
For the asymmetric mass uncertainties of Sgr A$^\star$ EHT and the
EHT+GRAVITY case, a two-sided Gaussian prior is used.
}}
\label{tab:priors}
\end{table}

\begin{table}[t]
\centering
\renewcommand{\arraystretch}{1.30}
\setlength{\tabcolsep}{5pt}

\resizebox{\textwidth}{!}{%
\begin{tabular}{lcccccccc}
\hline\hline

Dataset &
$\rho_{s,\max}$ &
$\rho_{s,\rm med}$ &
$\rho_s^{95\%}$ &
$r_{s,\rm med}$ &
$r_s^{95\%}$ &
$f_{\rm acc}$ &
$\tau_{\max}$ &
$N_{\rm step}/\tau_{\max}$ \\

&
\multicolumn{3}{c}{$[M_\odot\,{\rm pc}^{-3}]$} &
\multicolumn{2}{c}{$[{\rm kpc}]$} &
& & \\

\hline

M87$^\star$ EHT
& $\mathbf{25^\ast}$ & $\mathbf{1.93}$ & $\mathbf{11.83}$
& $\mathbf{16.41}$ & $\mathbf{54.36}$
& $\mathbf{0.410}$ & $\mathbf{265.2}$ & $\mathbf{377.0}$ \\

& 50  & 1.92 & 11.83 & 16.46 & 54.37 & 0.410 & 260.4 & 384.0 \\
& 100 & 1.91 & 11.84 & 16.46 & 54.34 & 0.410 & 266.5 & 375.3 \\
& 150 & 1.92 & 11.84 & 16.48 & 54.59 & 0.410 & 262.2 & 381.4 \\

\hline

Sgr A$^\star$ EHT
& 25  & 3.56 & 19.76 & 5.40 & 21.79 & 0.417 & 284.3 & 351.7 \\
& 50  & 4.20 & 29.43 & 5.10 & 21.22 & 0.405 & 322.1 & 310.5 \\

& $\mathbf{100^\ast}$ & $\mathbf{4.26}$ & $\mathbf{31.56}$
& $\mathbf{5.07}$ & $\mathbf{21.10}$
& $\mathbf{0.402}$ & $\mathbf{328.6}$ & $\mathbf{304.4}$ \\

& 150 & 4.27 & 31.59 & 5.07 & 21.02 & 0.402 & 321.2 & 311.4 \\

\hline

Sgr A$^\star$ GRAVITY
& 25  & 2.13 & 15.25 & 5.13 & 21.33 & 0.412 & 278.4 & 359.1 \\
& 50  & 2.18 & 16.60 & 5.09 & 21.17 & 0.410 & 270.7 & 369.4 \\

& $\mathbf{100^\ast}$ & $\mathbf{2.20}$ & $\mathbf{16.76}$
& $\mathbf{5.07}$ & $\mathbf{21.10}$
& $\mathbf{0.409}$ & $\mathbf{272.5}$ & $\mathbf{367.0}$ \\

& 150 & 2.22 & 16.66 & 5.06 & 21.05 & 0.410 & 283.0 & 353.4 \\

\hline

Sgr A$^\star$ EHT+GRAVITY
& 25  & 2.61 & 17.32 & 5.20 & 21.91 & 0.412 & 287.6 & 347.7 \\
& 50  & 2.81 & 21.09 & 5.07 & 21.20 & 0.408 & 304.0 & 329.0 \\

& $\mathbf{100^\ast}$ & $\mathbf{2.83}$ & $\mathbf{21.25}$
& $\mathbf{5.06}$ & $\mathbf{21.42}$
& $\mathbf{0.407}$ & $\mathbf{306.9}$ & $\mathbf{325.9}$ \\

& 150 & 2.83 & 21.20 & 5.06 & 20.81 & 0.409 & 288.5 & 346.6 \\

\hline\hline
\end{tabular}%
}

\caption{\textcolor{black}{
Posterior constraints, prior-sensitivity tests, and MCMC convergence
diagnostics for M87$^\star$ and Sgr A$^\star$.
The density prior is
$\rho_s\sim\mathcal{U}(0,\rho_{s,\max})$.
For M87$^\star$, $r_s\sim\mathcal{U}(10,80)$ kpc, while for
Sgr A$^\star$, $r_s\sim\mathcal{U}(3,40)$ kpc.
All runs use 128 walkers and $10^5$ steps, with the first 5000
steps discarded as burn-in and a thinning factor of 10.
Rows marked by an asterisk correspond to the priors adopted for
the final posterior analysis.
}}
\label{tab:mcmc_sensitivity}
\end{table}

\begin{table}[t]
\centering
\renewcommand{\arraystretch}{1.2}
\setlength{\tabcolsep}{5pt}

\resizebox{\textwidth}{!}{%
\begin{tabular}{lcccccc}
\hline\hline

Dataset &
$M$ &
$D$ &
$\rho_s$ &
$\rho_s^{95\%}$ &
$r_s$ &
$r_s^{95\%}$ \\

&
&
&
\multicolumn{2}{c}{$[M_\odot\,{\rm pc}^{-3}]$} &
\multicolumn{2}{c}{$[{\rm kpc}]$}
\\

\hline

M87$^\star$ EHT
&
$6.23^{+0.54}_{-0.59}\times10^9M_\odot$
&
$16.96^{+0.77}_{-0.76}$ Mpc
&
$1.93^{+4.97}_{-1.62}$
&
$<11.83$
&
$16.41^{+16.82}_{-5.05}$
&
$<54.36$
\\

Sgr A$^\star$ EHT
&
$3.63^{+0.48}_{-0.49}\times10^6M_\odot$
&
$8.16^{+0.15}_{-0.15}$ kpc
&
$4.26^{+12.98}_{-3.73}$
&
$<31.56$
&
$5.07^{+6.13}_{-1.64}$
&
$<21.10$
\\

Sgr A$^\star$ GRAVITY
&
$4.30^{+0.01}_{-0.01}\times10^6M_\odot$
&
$8.28^{+0.01}_{-0.01}$ kpc
&
$2.20^{+6.89}_{-1.93}$
&
$<16.76$
&
$5.07^{+6.16}_{-1.65}$
&
$<21.10$
\\

Sgr A$^\star$ EHT+GRAVITY
&
$4.02^{+0.33}_{-0.29}\times10^6M_\odot$
&
$8.22^{+0.08}_{-0.08}$ kpc
&
$2.83^{+8.71}_{-2.49}$
&
$<21.25$
&
$5.06^{+6.19}_{-1.63}$
&
$<21.42$
\\

\hline\hline
\end{tabular}%
}

\caption{\textcolor{black}{
MCMC constraints on the black-hole and dark-matter halo parameters.
The reported values are posterior medians with 68\% confidence levels,
together with one-sided 95\% upper limits on $\rho_s$ and $r_s$.
}}
\label{tab:posterior_constraints}
\end{table}

\textcolor{black}{
The choice of priors is important in the MCMC analysis, particularly in the presence of parameter degeneracies. Since the inferred constraints can depend on the adopted prior ranges, we select the priors carefully and test their influence on the posterior distributions.
}

\textcolor{black}{
For the BH mass $M$ and distance $D$, we adopt Gaussian priors centered on their measured values, with standard deviations given by the corresponding observational uncertainties,
\begin{equation}
\pi(\theta_i) \propto
\exp\left[
-\frac{1}{2}
\left(
\frac{\theta_i-\theta_{0,i}}{\sigma_i}
\right)^2
\right],
\end{equation}
where $\theta_i\in\{M,D\}$, $\theta_{0,i}$ is the corresponding measured value, and $\sigma_i$ denotes its observational uncertainty. When the uncertainty in $M$ is asymmetric, we adopt a two-sided Gaussian prior.
}

\textcolor{black}{
For the DM halo parameters $\rho_s$ and $r_s$, we adopt independent uniform priors. Since $\rho_s$ and $r_s$ show a strong degeneracy, the choice of their prior ranges can affect the inferred constraints. We therefore determine these ranges separately. For the scale radius $r_s$, we choose prior ranges based on previous observational studies of the corresponding galactic halos. In particular, we adopt
\begin{equation}
10 \leq r_s \leq 80~{\rm kpc}
\end{equation}
for M87$^\star$~\cite{Oldham_2018MNRAS,ChaoLi_2020MNRAS}, and
\begin{equation}
3 \leq r_s \leq 40~{\rm kpc}
\end{equation}
for Sgr~A$^\star$~\cite{Xiaowei_2024MNRAS, Cautun_2020MNRAS,Shen24PLB}.
}

\textcolor{black}{
For the characteristic density $\rho_s$, we test several upper bounds,
$\rho_{s,\max}=25$, $50$, $100$, and
$150~M_\odot\,{\rm pc}^{-3}$, while keeping the lower bound fixed at zero. We find that the posterior constraints for M87$^\star$ are already stable for
$\rho_{s,\max}=25~M_\odot\,{\rm pc}^{-3}$, whereas the results for Sgr~A$^\star$ become stable for
$\rho_{s,\max}\gtrsim100~M_\odot\,{\rm pc}^{-3}$.
Accordingly, we adopt
\begin{equation}
0\leq\rho_s\leq25~M_\odot\,{\rm pc}^{-3}
\end{equation}
for M87$^\star$ and
\begin{equation}
0\leq\rho_s\leq100~M_\odot\,{\rm pc}^{-3}
\end{equation}
for all Sgr~A$^\star$ datasets in the final MCMC analysis. The adopted priors are summarized in Table~\ref{tab:priors}.
}

The posterior probability distribution of the model parameters
$\Theta = (M, D, \rho_s, r_s)$ is obtained using Bayes' theorem,
which combines the likelihood function with the prior distributions.
It can be written as
\begin{equation}
P(\Theta | \mathcal{D}) =
\frac{\mathcal{L}(\mathcal{D} | \Theta)\,\pi(\Theta)}
{\mathcal{Z}},
\end{equation}
where $\mathcal{D}$ denotes the observational data, $\mathcal{L}$ is the likelihood function, $\pi(\Theta)$ represents the joint prior distribution of the parameters, and $\mathcal{Z}$ serves as a normalization
constant.

\textcolor{black}{
For each dataset, we use 128 walkers and run the MCMC chains for $10^5$ steps. The first 5000 steps are discarded as burn-in, and the remaining samples are thinned by a factor of 10. The convergence of the chains is checked using the mean acceptance fraction and the integrated autocorrelation time. The mean acceptance fractions are approximately $0.40$--$0.41$ for all datasets. The maximum integrated autocorrelation times range from about 265 to 329 steps, so that each chain contains more than 300 autocorrelation times. These values indicate that the chains are sufficiently long to sample the posterior distributions (see Table \ref{tab:mcmc_sensitivity}).
}

\textcolor{black}{
We also examine the dependence of the inferred halo constraints on the adopted prior ranges. For $\rho_s$, we repeat the analysis using upper prior bounds of $25$, $50$, $100$, and $150~M_\odot\,{\rm pc}^{-3}$. The constraints for M87$^\star$ remain nearly unchanged for $\rho_{s,\max}\geq25~M_\odot\,{\rm pc}^{-3}$, while those for Sgr~A$^\star$ become stable for $\rho_{s,\max}\gtrsim100~M_\odot\,{\rm pc}^{-3}$. For $r_s$, the 95\% upper limits obtained from the posterior distributions remain well below the adopted upper prior bounds. These results indicate that the constraints on $\rho_s$ are stable
against the tested prior ranges (see Table \ref{tab:mcmc_sensitivity}).
}

\begin{table}[t]
\centering
\begin{tabular}{lccc}
\hline\hline
Dataset &
$\chi^2_{\rm Schw,med}$ &
$\chi^2_{\rm halo,med}$ &
$\Delta\chi^2_{\rm med}$ \\
\hline

M87$^\star$ (EHT)
& 2.2310 & 0.3040 & 1.9270 \\

Sgr A$^\star$ (EHT)
& 0.5823 & 0.4047 & 0.1775 \\

Sgr A$^\star$ (GRAVITY)
& 0.4295 & 0.9511 & $-0.5216$ \\

Sgr A$^\star$ (EHT+GRAVITY)
& 0.2376 & 0.5553 & $-0.3177$ \\

\hline\hline
\end{tabular}
\caption{\textcolor{black}{
Comparison of the Schwarzschild and DM-halo predictions evaluated at the posterior median values. We define
$\Delta\chi^2_{\rm med}
=\chi^2_{\rm Schw,med}-\chi^2_{\rm halo,med}$,
such that positive values indicate a closer agreement of the DM-halo prediction with the observations, while negative values indicate a closer agreement of the Schwarzschild prediction.
}}
\label{tab:chi2_comparison}
\end{table}

\textcolor{black}{
Fig.~\ref{Fig.mcmc} shows the posterior distributions for Sgr~A$^\star$ and M87$^\star$. 
The shaded contours correspond to the 68\% and 95\% confidence levels, while the red dashed vertical lines indicate the 95\% upper limits on the DM halo parameters $\rho_s$ and $r_s$. 
The corresponding posterior median values and upper limits are summarized in Table~\ref{tab:posterior_constraints}. 
For M87$^\star$, we obtain the 95\% upper limits $\rho_s<11.83~M_\odot\,{\rm pc}^{-3}$ and $r_s<54.36~{\rm kpc}$. 
For Sgr~A$^\star$, the EHT, GRAVITY, and combined datasets give the 95\% upper limits $\rho_s<31.56$, $16.76$, and $21.25~M_\odot\,{\rm pc}^{-3}$, respectively. 
The corresponding posterior median values of $r_s$ are all close to $5~{\rm kpc}$, while the 95\% upper limits are around $21~{\rm kpc}$ for all three Sgr~A$^\star$ cases. 
Among the Sgr~A$^\star$ datasets, the GRAVITY-based analysis  provides the strongest constraint on $\rho_s$.
}

\begin{figure*}
    \centering
\includegraphics[scale=0.37]{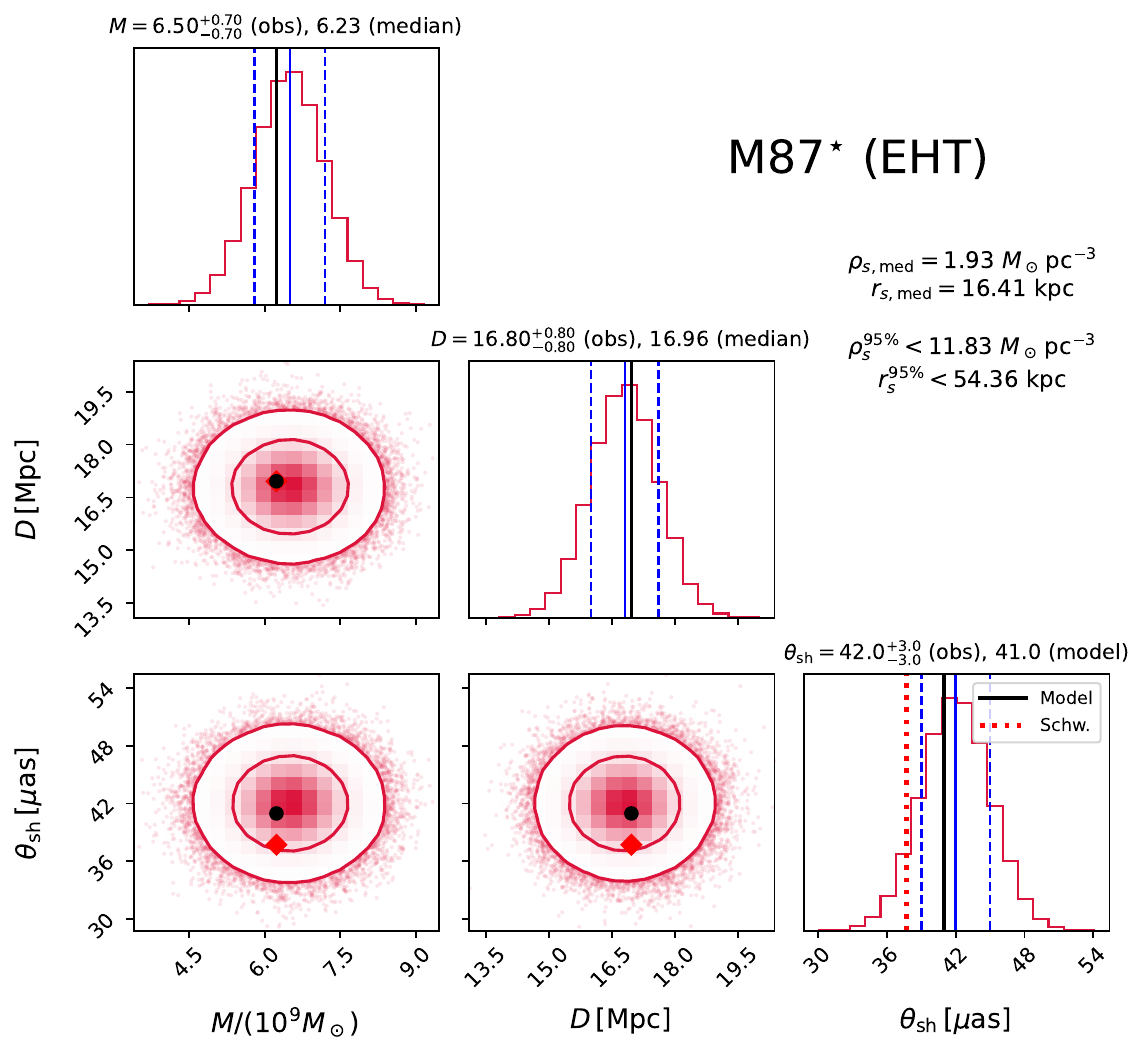}
\includegraphics[scale=0.37]{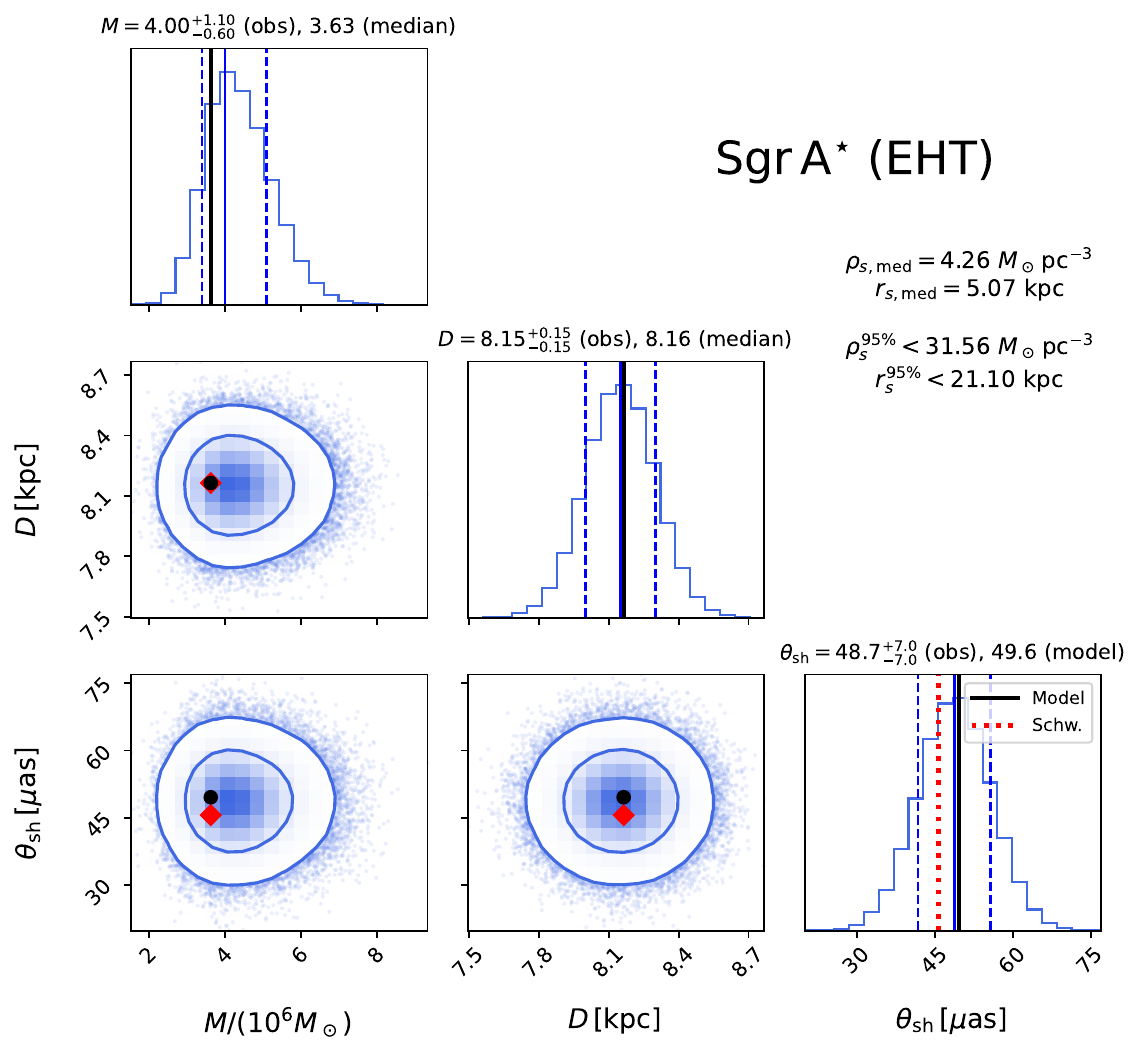}
\includegraphics[scale=0.37]{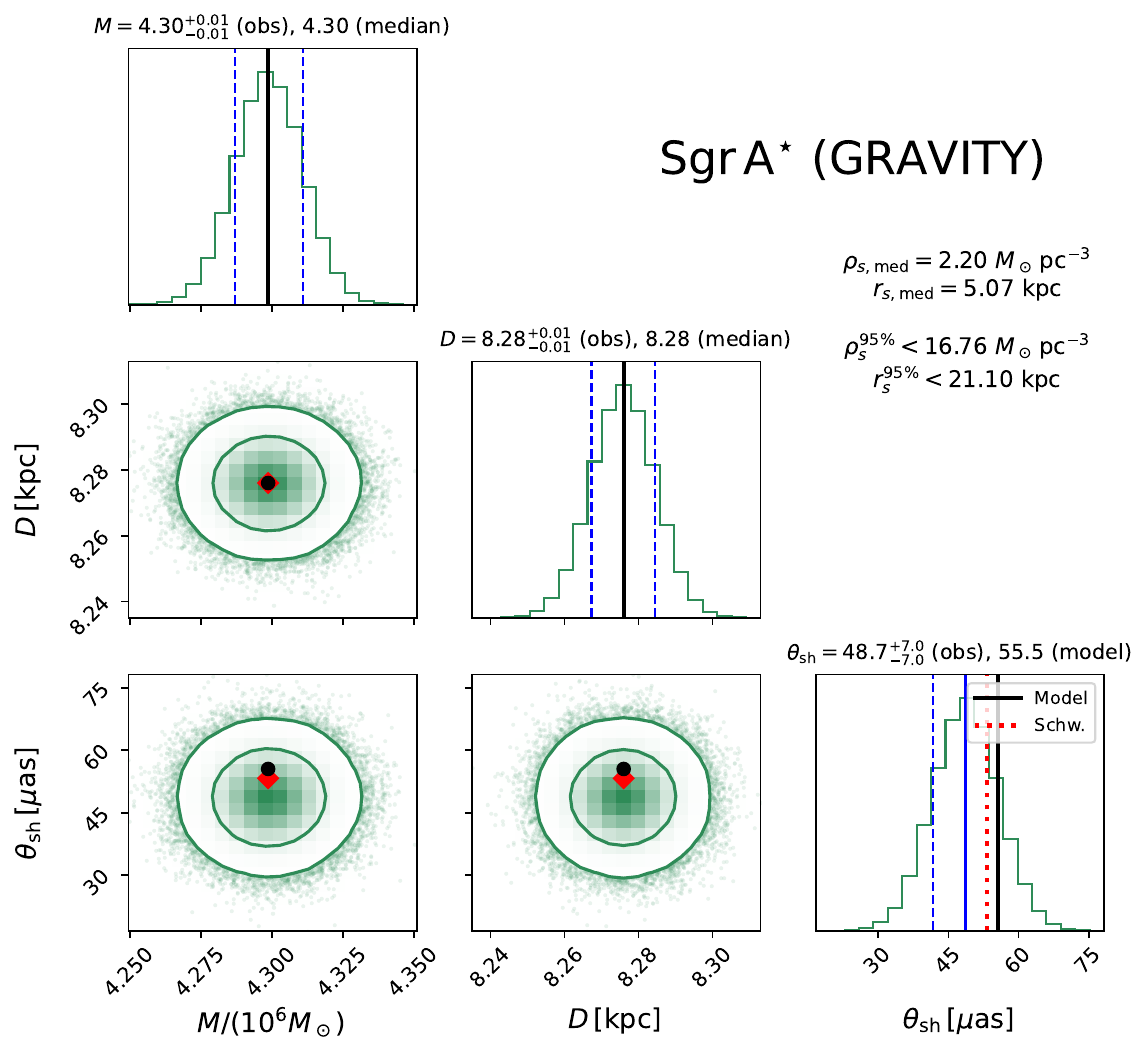}
\includegraphics[scale=0.37]{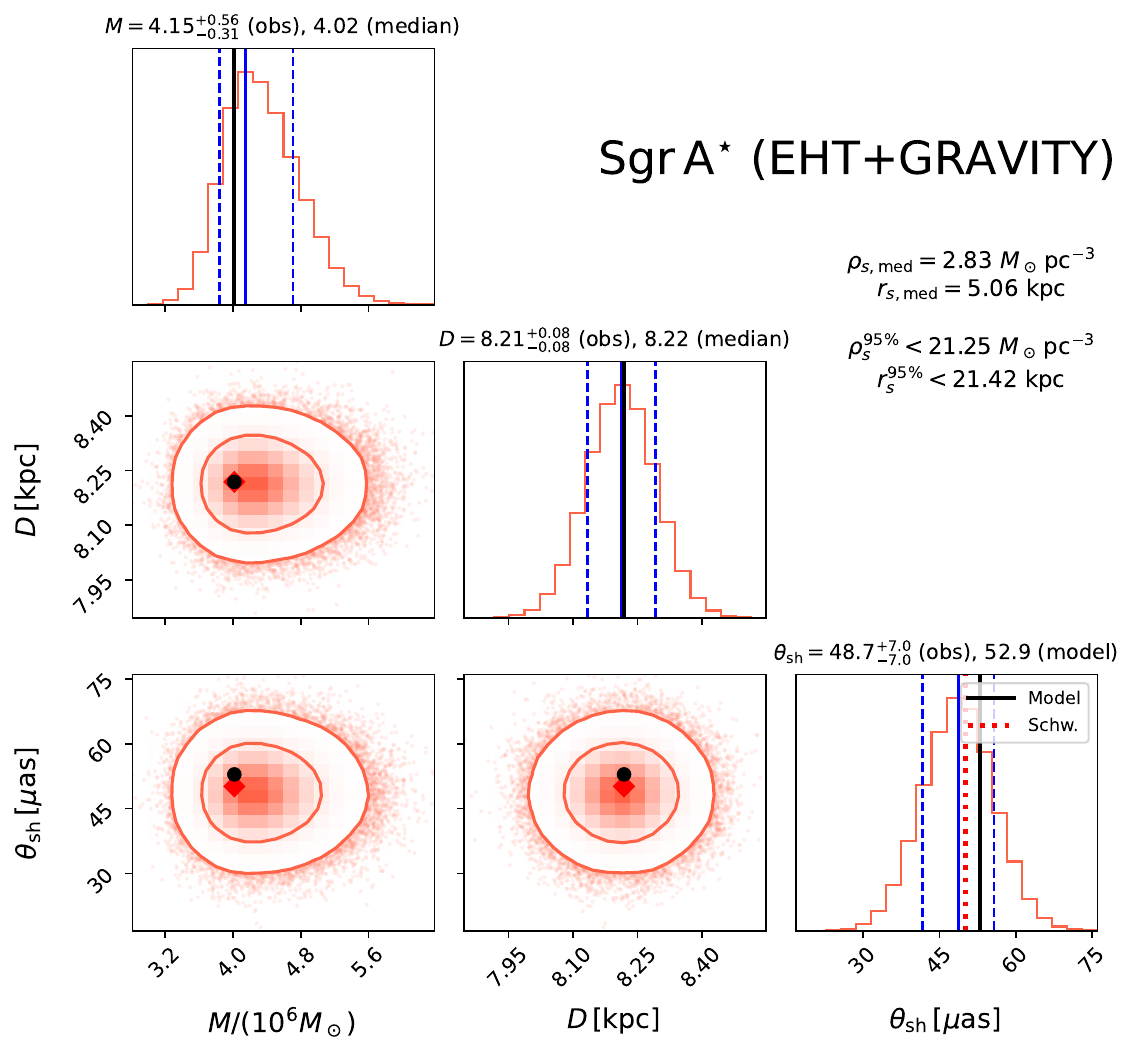}
    \caption{\textcolor{black}{ Comparison between the model predictions and observational data for the BH mass $M$, distance $D$, and shadow angular diameter $\theta_{\rm sh}$ for M87$^\star$ and Sgr~A$^\star$ using the EHT, GRAVITY, and combined (EHT+GRAVITY) datasets. The shaded regions show the observational distributions, while the solid and dashed blue vertical lines indicate the central observed values and their corresponding $1\sigma$ uncertainties, respectively. The black lines and circles mark the predictions obtained using the posterior median parameters of the DM halo model, while the red dotted lines and triangles show the corresponding Schwarzschild predictions. } }
    \label{Fig.comparison}
\end{figure*}

\textcolor{black}{
Fig.~\ref{Fig.comparison} compares the posterior-median predictions with the observational distributions of $M$, $D$, and $\theta_{\rm sh}$. For all datasets, the posterior-median values remain within the corresponding observational uncertainties. The EHT-based results for M87$^\star$ and Sgr~A$^\star$ show good agreement with the measured shadow diameters, while the GRAVITY and combined cases provide tighter constraints on the halo density parameter $\rho_s$. The upper limits on $r_s$ for the Sgr~A$^\star$ datasets remain very similar. The Schwarzschild predictions are also shown for comparison.
}

\textcolor{black}{
To quantify how closely the posterior-median predictions agree with the observations, we evaluate the total $\chi^2$ as
\begin{equation}
\chi^2 =
\left(\frac{M-M_{\rm obs}}{\sigma_M}\right)^2
+
\left(\frac{D-D_{\rm obs}}{\sigma_D}\right)^2
+
\left(\frac{\theta_{\rm sh}^{\rm th}-\theta_{\rm sh}^{\rm obs}}
{\sigma_{\theta_{\rm sh}}}\right)^2 .
\end{equation}
For the DM-halo model, $\chi^2_{\rm halo,med}$ is evaluated using the posterior median values \\ $(M_{\rm med}, D_{\rm med}, \rho_{s,\rm med}, r_{s,\rm med})$.
For the Schwarzschild case, $\chi^2_{\rm Schw,med}$ is calculated using the same $M_{\rm med}$ and $D_{\rm med}$, with $\rho_s=0$.
We define
\begin{equation}
\Delta\chi^2_{\rm med}
=
\chi^2_{\rm Schw,med}
-
\chi^2_{\rm halo,med}.
\end{equation}
As shown in Table~\ref{tab:chi2_comparison}, the DM-halo model gives a lower $\chi^2$ for the M87$^\star$ and Sgr~A$^\star$ EHT cases, although the improvement is small for Sgr~A$^\star$. In contrast, the Schwarzschild prediction gives a lower $\chi^2$ for the GRAVITY and combined EHT+GRAVITY cases. 
}

\section{Conclusion}\label{Sec:con}

Given that DM constitutes the dominant mass component of galaxies, its interaction with BHs at galactic centers can significantly influence the spacetime geometry in their vicinity. In particular, the presence of a DM halo may modify key observables such as orbital dynamics, gravitational lensing, and BH shadows. Therefore, investigating BH–DM systems not only deepens our understanding of BH environments but also provides a means to probe and constrain the properties and distribution of DM halo. Based on these considerations, we derived a new set of analytical Schwarzschild-like BH solutions that describe a static BH surrounded by a DM halo with a Dehnen-type density profile $(1,4,\gamma)$.

We investigated the properties of these BH–DM systems through their spacetime curvature, computing curvature invariants to characterize the singularity in the solution. Our analysis showed the presence of a spacetime singularity at $r=0$, as shown in Fig.~\ref{Fig.Curvature}. We further assessed the physical viability of the spacetime by examining the energy conditions associated with the DM halo. 
\cred{
The results showed that the NEC, WEC, and DEC are satisfied for all considered values of $\gamma$. The SEC is satisfied everywhere for $\gamma\geq2$, while for $\gamma<2$ it is violated only within the finite interval $0<r<r_{\rm SEC}$, with the exterior violation occurring only when $r_{\rm SEC}>r_h$. This behavior is illustrated in Fig.~\ref{Fig.EC}.
}

We further explored timelike and null geodesics in the BH–DM spacetime to examine how the derived analytical solutions influence the motion of massive and massless particles. From the analysis of the effective potential and the ISCO, we found that the ISCO radius $R_{\rm ISCO}$ increases with increasing density profile parameter $\gamma$ and DM scale radius $r_s$, indicating stronger gravitational effects. 
Moreover, increasing the DM halo parameters $r_s$ and $\rho_s$ shifted the unstable orbits outward, while the stable orbits moved inward (see Figs.~\ref{Fig.Veff}, \ref{Fig.Veff2}). Using observational data from the trajectories of Mercury and the S2 star, we determined the allowed parameter space $(r_s,\rho_s)$ for the chosen DM halo density profile $(\alpha,\beta,\gamma)=(1,4,9/4)$ and showed that its effects are more pronounced around supermassive BHs than in the solar system (see Fig.~\ref{fig:parspace}). The analysis of the shadow, photon sphere, and horizon structures showed that their radii increase with increasing $r_s$ and $\rho_s$, reflecting the enhanced gravitational influence due to the DM halo (see Fig.~\ref{Fig.shadow} and Table~\ref{tab:photonsphere}). 

\cred{
Finally, using observational data for M87$^\star$ and Sgr~A$^\star$, we constrained the characteristic density $\rho_s$ and scale radius $r_s$ of the DM halo through a Bayesian MCMC analysis. For M87$^\star$, we obtained the 95\% upper limits $\rho_s<11.83~M_\odot\,{\rm pc}^{-3}$ and $r_s<54.36~{\rm kpc}$. For Sgr~A$^\star$, the EHT, GRAVITY, and combined EHT+GRAVITY analyses yielded $\rho_s<31.56$, $16.76$, and $21.25~M_\odot\,{\rm pc}^{-3}$, respectively. In all three Sgr~A$^\star$ cases, the posterior median values of $r_s$ remain close to $5~{\rm kpc}$, with 95\% upper limits near $21~{\rm kpc}$. Among the Sgr~A$^\star$ datasets, the GRAVITY-based analysis therefore provides the tightest constraint on $\rho_s$. We also examined the dependence of the inferred halo constraints on the adopted priors and found that the constraints on $\rho_s$ remain stable against the tested upper prior bounds. A comparison based on the posterior-median predictions shows that the DM-halo model gives a lower $\chi^2$ for the M87$^\star$ and Sgr~A$^\star$ EHT cases, whereas the Schwarzschild prediction gives a lower $\chi^2$ for the GRAVITY and combined EHT+GRAVITY cases. 
}

Overall, our results for the exact BH-DM halo solutions highlight the role of DM halos in modifying the spacetime geometry and observable properties of astrophysical BHs, offering a new perspective on BH–DM interactions. Our findings demonstrate that observations of BH shadows and particle dynamics can be used to probe and constrain the properties of DM halo in the vicinity of supermassive BHs.

\acknowledgments{
\textcolor{black}{We are grateful to the anonymous referee for useful comments and suggestions that helped us improve the clarity and quality of the manuscript.}
}

%\vspace{10mm}
%%%%%%%%%%%Acknowledgements 
%\section*{Acknowledgements}

%\vspace{10mm}
%%%%%%%%%%%%%%%%%

%%%%%%%%%%%%%%%%%%bibliography
% \appendix
\bibliographystyle{apsrev4-1}  %% BibTeX style
\bibliography{mybib}  
\end{document}